\documentclass[%
reprint,
superscriptaddress,
amsmath,amssymb,
aps,longbibliography,
]{revtex4-2}

\usepackage{graphicx}%
\usepackage{verbatim}
\usepackage[version=4]{mhchem}
\usepackage{xcolor}
\usepackage{physics}
\usepackage[normalem]{ulem}
\usepackage{pdfpages}
\usepackage{pgffor}
\usepackage{booktabs}
\usepackage{bm}
\usepackage{overpic}
\usepackage{cleveref}
\usepackage{float} %
\usepackage{xspace}

\usepackage[normalem]{ulem}
\definecolor{tangerine}{rgb}{0.944,0.522,0}
\definecolor{verde}{rgb}{0.,0.6,0}
\definecolor{rosso}{rgb}{0.9,0.0,0.2}
\definecolor{orange}{rgb}{1.0,0.5,0.0}

\newif\ifhighlight

\newcommand{\highlight}{\highlighttrue}
\highlight
\newcommand{\editor}[2]{%
  \expandafter\newcommand\csname #1note\endcsname[1]{%
    \textcolor{#2}{(\textbf{#1note:} \textsc{##1})}}%
  \expandafter\newcommand\csname #1\endcsname[1]{%
    \ifhighlight\textcolor{#2}{##1} \else ##1\fi}%
  \expandafter\newcommand\csname #1cancel\endcsname[1]{%
    \ifhighlight\textcolor{#2}{\sout{##1}}\fi}%
  \expandafter\newcommand\csname #1change\endcsname[2]{%
    \ifhighlight\textcolor{#2}{\sout{##1} ##2}\else ##2\fi}%
  \newenvironment{#1text}{\ifhighlight\color{#2}\fi}{\color{black}}
}

\editor{MK}{orange}
\editor{TEIH}{blue}
\newcommand{\grayout}[1]{}

\usepackage{xspace}
\newcommand{\uaml}{ML-UAFD\xspace}
\newcommand{\uafd}{UAFD\xspace}
\newcommand{\petuafd}{PET-UAFD\xspace}
\newcommand{\petuafdh}{PET-EXP\xspace}

 \newcommand{\makebm}[1]{\expandafter\def\csname bm#1\endcsname{\bm{#1}}}

\def\makebmloop#1{\ifx#1\relax\else\makebm{#1}\expandafter\makebmloop\fi}
\makebmloop ABCDEFGHIJKLMNOPQRSTUVWXYZabcdefghijklmnopqrstuvwxyz\relax

\newcommand{\dm}{\bm{\Phi}}
\newcommand{\mlip}[1]{V_{#1}^{\text{ML}}}
\newcommand{\gdist}{\mathcal{N}}

\newcommand{\rscan}{\text{r$^2$SCAN}}

\makeatletter
\AtBeginDocument{\let\LS@rot\@undefined}
\makeatother

\begin{document}

\title{
Errors that matter: Uncertainty-aware \\ universal machine-learning potentials calibrated on experiments}

\author{Matthias Kellner}\thanks{These authors contributed equally to this work.}
\affiliation{Laboratory of Computational Science and Modeling, Institut des Mat\'eriaux, \'Ecole Polytechnique F\'ed\'erale de Lausanne, 1015 Lausanne, Switzerland}

\author{Teitur Hansen}\thanks{These authors contributed equally to this work.}
\affiliation{Department of Physics, Technical University of Denmark}

\author{Thomas Bligaard}
\affiliation{Department of Energy Conversion and Storage, Technical University of Denmark}

\author{Karsten Wedel Jacobsen}
\email{kwja@dtu.dk}
\affiliation{Department of Physics, Technical University of Denmark}

\author{Michele Ceriotti}
\email{michele.ceriotti@epfl.ch}
\affiliation{Laboratory of Computational Science and Modeling, Institut des Mat\'eriaux, \'Ecole Polytechnique F\'ed\'erale de Lausanne, 1015 Lausanne, Switzerland}

\newcommand{\CSM}[1]{{\color{red}#1}}

\date{\today}%

\begin{abstract}
Machine-learning (ML) models of atomic-scale interactions achieve the accuracy of the quantum mechanical calculations on which they are trained, but at a dramatically lower computational cost. Their predictions can be made trustworthy by uncertainty quantification techniques that estimate the residual error relative to their reference. 
These errors, however, do not include uncertainty contributions from the approximations inherent in the electronic structure calculations, which are often the main source of discrepancy with empirical observations.
We construct an ensemble of ML potentials trained on multiple electronic-structure references and calibrate it against experimental data on cohesive energies, atomization energies, lattice constants and bulk moduli of simple materials and molecules, similar to the uncertainty-aware functional distribution (UAFD) approach.
The resulting ensemble of models, which we call PET-UAFD, can be used to simulate matter across a wide range of compositions and thermodynamic conditions.
By comparison with experimental measurements of the density and structure of liquids, we demonstrate that, even outside the static properties on which it was calibrated, PET-UAFD enables predictions that are as accurate against experiments as the best available electronic-structure reference and that the spread in the ensemble can be used to assess the reliability of such predictions.
We also introduce the \petuafdh protocol that uses shallow ensembles and statistical reweighting techniques to provide accurate estimates of uncertainty relative to experimental measurements at virtually no additional cost over a simulation based on a single conventional ML potential. 
Ultimately, this approach provides a practical and inexpensive approach to elevate machine-learning potentials from faithful interpolators of approximate theories to genuinely predictive tools anchored in experimental reality.

\end{abstract}

\maketitle

\section{Introduction}

Simulating matter at the atomic scale, based on an explicit description of the quantum mechanical behavior of electrons and nuclei, has long held the promise of facilitating the interpretation of experiments and accelerating the discovery of materials and molecules with novel or improved properties~\cite{hafnerComputationalMaterialsDesign2006}.
Exact modeling of quantum mechanics in many-body problems has a steep computational cost, which has led to the development of approximate techniques -- with the many flavors of density-functional theory (DFT) often achieving a good balance between effort and accuracy~\cite{burk12jcp}. 
The rise of models that approximate quantum mechanics using machine-learning techniques -- above all, machine-learning interatomic potentials (MLIPs) that predict energy and forces given the position of the atoms -- has provided an even better balance, pushing further the length and time scales accessible to atomistic modeling with the accuracy of quantum mechanics.

As MLIPs become more transferable and ``universal'' models that can handle arbitrary structures and compositions are developed~\cite{chenUniversalGraphDeep2022, dengCHGNetPretrainedUniversal2023, bata+25jcp, woodUMAFamilyUniversal2026, mazi+25ncomm}, it is essential to assess whether their predictions for a given system can be trusted. 
This need has triggered the development of uncertainty-quantification techniques~\cite{musi+19jctc,imba+21jcp,pern22jcp,kell-ceri24mlst,bigi+24mlst,zhuFastUncertaintyEstimates2023,perezUncertaintyQuantificationMisspecified2025}, which predict how closely the ML model matches the electronic-structure reference it is trained on.
These techniques, however, do not consider that approximate electronic-structure calculations themselves often are far from a perfect proxy for physical reality, with different flavors of DFT achieving varying degrees of success depending on the problem at hand. 

The availability of a multitude of DFT exchange-correlation approximations has been exploited in the past to build ``DFT ensembles'' \cite{ mortensenBayesianErrorEstimation2005, wellendorffDensityFunctionalsSurface2012a, 
aldegundeDevelopmentExchangeCorrelation2016,
wellendorffMBEEFAccurateSemilocal2014a, hansenUncertaintyawareElectronicDensityfunctional2025}, calibrated against empirical observations to provide estimates of the error that actually matters: the error against experiments. 
In this work, we demonstrate that, by combining DFT ensembles with uncertainty-aware MLIPs, it is possible to reach the ultimate goal of a universal approximation of the interatomic potential that reaches an accuracy approaching that of established electronic structure methods and provides error estimates on complex thermodynamic properties that reflect their accuracy against physical reality.

\section{Theory}

Our framework leverages recently developed techniques to generate distributions of DFT functionals and calibrate them against experiments on the structural and energetic properties of simple molecules and solids (the uncertainty-aware functional distribution (UAFD) framework~\cite{hansenUncertaintyawareElectronicDensityfunctional2025}). 
By combining these ideas with expressive ML potentials trained against DFT references~\cite{pozd-ceri23nips,bigiPushingLimitsUnconstrained2026} we make the evaluation of more complicated thermodynamic properties affordable, such as the density and structure of liquids. 
By further implementing a direct propagation of shallow ensembles (DPOSE) framework, we make the computational overhead relative to simulations performed with a single DFT-trained MLIP negligible. 
In this section, we briefly summarize these ideas, highlighting the implementation changes that need to be made to fully exploit their synergy.

\subsection{Uncertainty-aware MLIP distributions}

We base our framework for estimating the uncertainty of universal MLIP (uMLIP) predictions against experiment on the uncertainty-aware functional distribution (UAFD) framework~\cite{hansenUncertaintyawareElectronicDensityfunctional2025}. In the UAFD, the variations between predictions with different exchange-correlation (xc) functionals are used to estimate errors in predicted quantities. We use the same idea here, but instead of working with the DFT calculations themselves, we work directly with a model space of MLIPs fitted to different xc-functionals. To be specific, we consider a linear space spanned by a basis set of MLIPs ($\mathcal{B}=\{\mlip{\text{PBE}},\mlip{\text{PBEsol}}, \mlip{\text{PBE+U}}, \mlip{\rscan}, \mlip{\text{\rscan-D3}}\}$) trained on the indicated exchange-correlation functionals. 
For a given atomic system, $A$, a property $y$ can be predicted as $y(A|\bmw)= \sum_i w_i V_i^{\text{ML}}(A)$, where the sum runs over the basis set, and $\bmw$ is a vector determining the linear combination of the MLIP basis. We use the sum rule $\sum_i w_i = 1$, which ensures the right behavior if the zero-point of the potentials is shifted, leading to $m=4$ effective parameters.

Following Ref.~\onlinecite{hansenUncertaintyawareElectronicDensityfunctional2025}, we define a normal distribution over the model space $\mathcal{P}_\mathcal{M}(\bmw)=\mathcal{N}(\bmw|\bmw_0,\bmK)$, with parameter mean $\bmw_0$ and covariance matrix $\bmK$. To determine the parameter mean and the covariance matrix, we introduce a database of $N$ systems, $A_n, (n = 1,2, \ldots N)$, with target values $t_n$ for the considered property as obtained either from experiments or high-fidelity calculations. The predictions of the data points by the different MLIP models are collected in the design matrix $\dm_{ni} = \mlip{i}(A_n)$. The probability of reproducing a data point is $p(t_n) = \int \delta(t_n-y(A_n|\bmw)) \mathcal{P}_\mathcal{M}(\bmw)\,d\bmw$.
The parameter mean and covariance matrix may now be obtained by maximizing the product of these probabilities over all data points. Using a Gaussian ansatz for $\mathcal{P}_\mathcal{M}$, this corresponds to minimizing the negative log-likelihood cost function \cite{hansenUncertaintyawareElectronicDensityfunctional2025}
\begin{equation}\label{eq:loglike}
\mathcal{C}(\bm{w}_0,\bm{K}) = \sum_n (t_n-\bar{y}_n)^2/\sigma_n^2+ \sum_n \log(\sigma_n^2) + N\log(2\pi),
\end{equation}
where we have introduced the mean predictions $\bar{y}_n = (\dm\bmw_0)_n$ and the variances $\sigma_n^2 = (\dm\bmK\dm^T)_{nn}$. We add regularization terms to this cost function in the same way and with the same parameter values as in Ref.~\onlinecite{hansenUncertaintyawareElectronicDensityfunctional2025}.

Having determined $\bmw_0$ and $\bmK$, the resulting distribution $\mathcal{P}_\mathcal{M}(\bmw)=\mathcal{N}(\bmw|\bmw_0,\bmK)$ can be used to make a predictive distribution of, in principle, any new property $z(A|\bmw)$, which can be calculated from the MLIPs in the basis. The predictive distribution is given in the same way as for a data point as $p(z) = \int \delta(z-z(A|\bmw)) \mathcal{P}_\mathcal{M}(\bmw)\,d\bmw$. If the quantity depends linearly on the parameters, the evaluation of the mean and variance of the prediction can be performed analytically, but we shall go beyond this here and make an estimate based on a minimal ensemble. We construct the ensemble by introducing new variables $\tilde{\bmw} = \bmD^{-1/2}\bmU(\bmw-\bmw_0)$, where $\bmU$ is the unitary matrix diagonalizing $\bmK$: $\bmD = \text{diag}(k_1, k_2, \ldots k_m)=\bmU\bmK\bmU^T$, with $k_i$ being the eigenvalues of $\bmK$.
In order to calculate, for example, the variance, we rewrite
\begin{align}\label{eq:sampling}
    &\langle \left(z-\langle z\rangle\right)^2\rangle 
     =     \int (z(A|\bmw)-\langle z\rangle)^2 \gdist(\bmw|\bmw_0,\bmK)\,d\bmw\\ \nonumber
    & = \int \left[z\left(A|\bmw_0+\bmU^T\bmD^{1/2}\tilde{\bmw}\right)-\langle z\rangle\right]^2\,
    \gdist(\tilde{\bmw}|0,\bm{1})\,d\tilde{\bmw}\\ \nonumber
    & \approx \sum_{\alpha=1}^m \left[z\left(A|\bmw_0+\bmU^T\bmD^{1/2}\bme_\alpha\right) -z(A|\bmw_0)       \right]^2,\nonumber
\end{align}

where in the last step, we sample the standard normal distribution at the points $\bme_\alpha$, which are unit vectors pointing in the $\alpha$th direction. It is straightforward to show that the expression for the variance Eq.~(\ref{eq:sampling}) is correct up to second-order variations of $z$ with the parameters.

Calculating the mean and variance of $z$ therefore requires the evaluation of $m+1$ models corresponding to $\bmw=\bmw_0$ and 
\begin{equation}
    \bmw_\alpha = \bmw_0 + \bmU^T\bmD^{1/2}\bme_\alpha = \bmw_0 + \sqrt{k_\alpha}\bmu_\alpha,\  \alpha=1\ldots m,
\end{equation}
where $\bmu_\alpha$ denotes the $\alpha$th eigenvector of $\bmK$.
This formulation would be prohibitively expensive if one wanted to apply it using explicit DFT calculations, since one needs to run one simulation for each of the $m+1$ potentials, and each potential $V_\alpha=\sum_i w_{\alpha i} V^{\text{DFT}}_i$ requires the evaluation of $m$ electronic-structure calculations.
However, if each of the DFT functionals is approximated by a separate MLIP, $V^{\text{ML}}_i$, the cost becomes affordable. 
The UAFD framework can then be applied using the potentials, corresponding to an uncertainty-aware MLIP distribution.
We will perform most of the demonstrative examples using independent MLIPs and separate trajectories, to demonstrate the idea without introducing additional approximations.
It is, however, possible to eliminate the overhead almost entirely, using ideas from Ref.~\citenum{kell-ceri24mlst}. 
First, we can tune a multi-head potential that shares most of the computationally demanding parts of the architecture, and then predict different DFT targets using separate heads. As we will show, this entails a small increase in the model error, but makes the evaluation of each UAFD combination $V_\alpha$ only marginally more expensive than computing a single DFT-trained MLIP. 
Second, in many cases, it is possible to use statistical reweighting~\cite{imba+21jcp} to compute observables for each $V_\alpha$ by running a single trajectory using $V_0$ and then correcting the distribution to produce the same average as a simulation run with each of the members of the ensemble.

\subsection{Inexpensive uncertainty propagation using direct propagation of shallow ensembles}

\uafd predictions of experimental quantities, such as atomization energies, can readily be computed from the direct MLIP potential energy predictions.
Simple transformations of the output $y$ of a model for a single structure $A$, $z(A)=f(y(A))$, can also be obtained easily by linearizing the transformation, which can also be extended quite easily to quantities that can be written as combinations of a few single-point calculations (e.g. cohesive energies, defect energies, and elastic constants). 
Quantities such as finite-temperature thermodynamic averages that need more complicated simulation workflows for modeling require a different approach.
Separate simulations with different models should be performed, computing the value of the observable and its uncertainty as the mean and the standard deviation of the values $\{z_\alpha\}$ obtained from the various simulations. 
In the case of UAFD, where the ensemble is built with a minimum number of members characterized by their weights $\bm{w}_\alpha$, the expression that matches the error estimate for a Gaussian distribution in the linear case is given by 
\begin{equation}
\sigma^2_{z} = \sum_{\alpha=1}^{m}(\expval{z}_{\bm{w}_\alpha}-\expval{z}_{\bm{w}_0})^2\label{eq:uafd-sigmaz},
\end{equation}
where $\expval{z}_{\bm{w}_\alpha}$ indicates the average of the target observable computed by sampling configurations using the potential $V_\alpha$ corresponding to the weights ${\bm{w}_\alpha}$.
Evaluating~\eqref{eq:uafd-sigmaz} entails an overhead of a factor of $(m+1)^2$ over a simulation with a single MLIP, because each $\expval{z}_{\bm{w}_\alpha}$ requires a separate simulation and because evaluating each member of the potential ensemble requires computing all $(m+1)$ MLIPs trained on the basis functionals. 

In order to eliminate this overhead, we apply the direct propagation of shallow ensembles (DPOSE) scheme~\cite{kell-ceri24mlst}. 
First, the ensemble of DFT functionals can be computed as a \emph{shallow ensemble}, i.e. with an architecture in which the members of the ensemble share the computationally intensive backbone, and differ only in how the backbone features are
combined to predict energy and forces. 
This approach has been demonstrated to produce error estimates as accurate as ensembles of independent models~\cite{kell-ceri24mlst}, and allows the use of a large number of ensemble members and the more flexible optimization of the distribution, propagating information of the uncertainties in the hidden layers of the MLIP neural networks~\cite{schaferHowTrainShallow2026}. 
Here, for simplicity, we still train models on the DFT functionals (as discussed further in Section~\ref{sec:petuafd_training}), using separate heads for the different functionals, and compute the ensemble members using optimized weights, and the ``best'' potential is that obtained from the $\bm{w}_0$ weights.

Computing~\eqref{eq:uafd-sigmaz} using this shallow potential would entail an overhead of approximately $(m+1)$, because one still needs to perform separate simulations to sample the members of the ensemble.
This residual overhead can be eliminated by using statistical reweighting~\cite{torr-vall99jcp} to compute averages consistent with each potential using a single trajectory.
In practice, we use a more stable, approximate, cumulant expansion, following Refs.~\cite{ceri+12prsa, imba+21jcp}:
\begin{equation}
\label{eq:reweight-cea}
\left<z\right>_{\bm{w}_\alpha} \approx \expval{z}_{\bm{w}_0} - \beta [\expval{z(V_\alpha-V_0)}_{\bm{w}_0} - \expval{z}_{\bm{w}_0}\expval{V_\alpha-V_0}_{\bm{w}_0}].
\end{equation}
Conveniently, in Eq.~\ref{eq:reweight-cea}, any constant shift of the potential energy predictions $V_\alpha$ cancels out exactly, removing the need to normalize the individual MLIP-DFT predictions that may be shifted due to the approximations and implementation details in the reference DFT computations.
Overall, combining the UAFD scheme with the DPOSE framework allows one to evaluate uncertainties against experiments for complex thermodynamic properties at the same cost as a simulation using a single conventional MLIP.

\subsection{Machine-learning interatomic potentials}

We use machine-learning interatomic potentials (MLIPs) to efficiently approximate the basis functionals in the \uafd framework. The MLIP serves as a surrogate model that approximates the computation of the potential energy $V$ of the base functionals.
In order to ensure transferability and computationally efficient inference, MLIPs employ an atom-centered approach to approximate the total potential energy $V$ of a structure $A$, as the sum of atom-centered contributions $v(A_i)$ that are computed from local atomic environments $A_i$ around the central atom $i$. 
\begin{equation}
V(A) = \sum_{i=1}^{N_{\text{atoms}}} v(A_i)
    \label{eq:per_atom}
\end{equation}
The domain of applicability of the MLIP is determined by the composition of the training data, and the flexibility of the functional form chosen to parameterize $v$. 
Transferable MLIPs that are not restricted to one material class can be obtained by training on structurally diverse databases of reference DFT calculations that adequately cover material space. 
We discuss the specific choices of architecture and hyperparameters that we use here in Sec.~\ref{sec:methods}. 
There is, however, no specific limitation that would prevent applying the same combination of UAFD and DPOSE that we use here to any other MLIP framework.

\section{Methods}\label{sec:methods}

\subsection{Training of the \uafd basis-function MLIPs }\label{sec:petuafd_training}
In this work, we introduce an ensemble of machine-learned universal DFT approximators, also commonly referred to as universal machine-learning force fields or uMLIPs.
This ensemble of DFT approximators serves as the basis for the \uafd.
The universal MLIP approximators are based on the point edge transformer architecture from Pozdnyakov et al. \cite{pozd-ceri23nips}, using the implementation in Ref.~\citenum{bigiPushingLimitsUnconstrained2026} with hyperparameters corresponding to PET-OAM-L models using a fixed local cutoff radius of 4.5~\AA. 
In this work, we introduce two uMLIP ensembles that differ by the degree of weight sharing: The first ensemble, denoted \textbf{\petuafd}, is an ensemble of independent models in which each model serves as a surrogate model for one reference DFT level of theory.
The second ensemble, which forms the basis of the \textbf{\petuafdh} protocol, also predicts an ensemble of DFT labels; however, most of the weights are shared between the ensemble models and only the final read-out heads differ.
The additional read-out heads only account for about 16 $\%$ of the total model parameters. The computational overhead of obtaining an ensemble of energy predictions is therefore greatly reduced compared to \petuafd. 
Given the tremendous cost of  large-scale DFT computation campaigns that are required to generate sufficiently diverse datasets for the training of universal MLIPs, we restrict ourselves to existing datasets to demonstrate the applicability of our method. This effectively constrains the target xc functionals on which we can train uMLIPs to the PBE family, namely PBE \cite{perdewGeneralizedGradientApproximation1996b} from the MATPES dataset \cite{kaplanFoundationalPotentialEnergy2025a}, PBEsol \cite{perdewRestoringDensityGradientExpansion2008a} from the MAD dataset \cite{mazi+25sd}, and PBE+U from the OMAT24 \cite{barroso-luqueOpenMaterials20242024} database. Additionally, we train one MLIP targeting \rscan\  \cite{furnessAccurateNumericallyEfficient2020} as the target level of theory using the MATPES database. Finally, we add one MLIP to the basis functions that also targets the dispersion corrected \rscan-D3 level of theory \cite{grimmeConsistentAccurateInitio2010a}. The training details of the uMLIPs are given in the SI section~\ref{sec:SI_training_details} and details about the reference training sets in section~\ref{sec:training_databases}. 

\subsection{Calibration of the \uaml on experimental reference data: Generating PET-UAFD}

We train a UAFD using four databases containing atomization energies, cohesive energies, lattice constants, and bulk moduli. These are the same databases used in Ref.~\onlinecite{hansenUncertaintyawareElectronicDensityfunctional2025}. There are 222 molecules whose target values for atomization energies are from G3/99\cite{curtissAssessmentGaussian3Density2000}, and 44 solids whose target values for the cohesive energies, lattice constants and bulk moduli come from Tran et al. \cite{tranRungs142016}. All systems are geometrically optimized using the MLIPs in the basis, and the properties are calculated subsequently. We used 5-fold cross-validation to test the method throughout this process and identified some data points as outliers. Some of the outliers appear because we are far outside the training regime of the MLIPs. For example, the cutoff radius of the MLIPs is so small that the noble gas solids do not form. Other outliers are less obvious (e.g. incorrect spin state for the isolated atoms in the training set of the MLIP) but in any case they are removed to obtain a robust fit. A detailed description of which points are removed can be found in the SI section~\ref{sec:si_uafd_outlier_removal}.
Following the pruning of the dataset, we train a single UAFD based on all the remaining data points, using the theory described above to create the tailored ensemble.

\subsection{Experimental database of liquid metal densities and radial distribution functions}
\label{sec:database_substances}
We evaluate the quality of \petuafd and \petuafdh uncertainty estimates and predictions of liquid densities and radial distribution functions by comparing them with the corresponding experimental data. For this purpose, we assemble a database of 21 liquid systems, including elemental
liquids and compounds, namely Na, Bi, Sc, Si, Ti, NaCl, H$_2$O, Ge, V, Ga, Sn, Ni, Co, Cu, Zn, GaAs, Sb, Fe, As, Mn, and Cr. Most liquids were measured near the respective melting
points of the substances -- except for a few entries for which we could also retrieve experimental measurements for liquids well above their melting points. The database is gathered from various sources in the literature. A detailed listing of the references for individual conditions is given in the Supporting Information section~\ref{sec:experimental_references}, Table~\ref{tab:si_list_experimental_references}.

\subsection{Molecular dynamics simulations with \petuafd}
We compute ensemble averages of densities and radial distribution functions in the NPT ensemble under the respective experimental conditions. We used an integration time step of 0.5~fs for liquid water, 2~fs for Na, NaCl, Si, Ge and first-row transition metals, and 4~fs for heavier elements Ga, Sn, Bi, As, Sb, and the III-V semiconductor GaAs. For optimal sampling, we combine a local stochastic velocity rescaling thermostat \cite{buss+07jcp} employing a coupling time $\tau$ of 10~fs with a colored noise thermostat \cite{ceri+10jctc}. Barostatting is performed with a modified isotropic Bussi, Zykova-Timan, Parrinello barostat \cite{buss+09jcp}  with a coupling time of 500~fs, and using a GLE colored-noise thermostat to keep the cell velocity distribution at the target temperature \cite{ceri+14cpc}.
All molecular dynamics simulations were performed using i-PI version~3 \cite{litm+24jcp} and the \texttt{metatomic} interface that implements the infrastructure needed to apply the DPOSE method \cite{bigi+26jcp}. The initial molten structures of all elements and compounds studied in this work were obtained by melting cubic boxes of the corresponding solid using short molecular dynamics simulations at three times the experimental melting point. Input structures were constructed to contain about 500 atoms. All simulations were performed for at least $10^5$ time steps. Radial distribution functions were computed using the MDAnalysis code \cite{gowersMDAnalysisPythonPackage2016, michaud-agrawalMDAnalysisToolkitAnalysis2011}.

\subsection{Accelerated \uafd simulations with \petuafdh}

Simulations were performed to assess the reliability of the accelerated error estimation protocol with the multi-head \petuafdh model, using MD simulation parameters identical to those for \petuafd. Reweighted RDFs for the respective members were obtained via CEA reweighting (eq.~\ref{eq:reweight-cea}) from the trajectory generated with the fine-tuned $\bm{w}_{0}$ \petuafdh ensemble member, using energies computed for simulation snapshots written with a sampling interval of 100 time steps. 
We reused the covariance matrix and the weights of the \petuafd ensemble members to generate head-\uafd ensemble members, which underscores the robustness of the procedure to determine $\bm{w}_\alpha$, which is not significantly affected by the small discrepancies between the predictions of the explicit ensemble and its shallow counterpart.

\section{Results}

\subsection{Evaluation of MLIPs and \petuafd on \uaml training data}

We begin by considering the behavior of the base MLIPs $\mathcal{B}=\{$$\mlip{\text{PBE}}$, $\mlip{\text{PBEsol}}$, $\mlip{\text{PBE+U}}$, $\mlip{\rscan}$, $\mlip{\text{\rscan-D3}}\}$ on the four (pruned) datasets of molecular atomization energies, solid cohesive energies, lattice constants, and bulk moduli. The root-mean-square errors (RMSE) are shown in Table~\ref{tab:UAFD_train_MLIP_RMSE_eval}. For comparison, the errors obtained from DFT calculations with the PBE functional are also shown. (The PBE values are from Ref.~\onlinecite{hansenUncertaintyawareElectronicDensityfunctional2025}). Comparing the results of $\mlip{\text{PBE}}$ and DFT-PBE, it is clear that the errors on the atomization energies and the cohesive energies are much larger with $\mlip{PBE}$ than with DFT-PBE, while they are rather similar for the lattice constants and the bulk moduli. 
This is to be expected, as the atomization energies and cohesive energies represent much larger energy differences also involving the energies of single atoms. Since most of the MLIP training data focus on perturbations of bulk structures, the $\mlip{\text{PBE}}$ predictions of lattice constants and bulk moduli are much closer to those obtained with DFT-PBE.

\begin{table}[ht]
    \centering
    \caption{RMSE evaluation on atomization energies (AE), cohesive energies (CE), lattice constants (LC) and bulk moduli (BM) for different MLIPs.}
    \label{tab:UAFD_train_MLIP_RMSE_eval}
    \resizebox{0.38\textwidth}{!}
    {\begin{tabular}{l c c c c}
        \toprule
                & AE & CE & LC & BM\\
         Models & [eV/atom] & [eV/atom] & [Å] & [GPa] \\
        \midrule
        $\mlip{\text{PBE}}$     & 0.37 & 0.53 & 0.08 & 18.4 \\
        $\mlip{\text{PBE+U}}$ & 1.06 & 1.51 & 0.08 & 20.9\\
        $\mlip{\text{PBEsol}}$ & 1.36 & 1.86 & 0.03 & 24.6 \\
        $\mlip{\text{\rscan}}$ & 0.43 & 0.53 & 0.03 & 11.5 \\
        $\mlip{\text{\rscan-D3}}$  & 0.33 & 0.54 & 0.03 & 11.7 \\
        $\bm{w}_0$      & 0.29 & 0.57 & 0.03 & 11.5 \\
        \midrule
        DFT-PBE  & 0.18 & 0.26 & 0.08 & 16.5 \\
        \bottomrule
    \end{tabular}}
\end{table}

The overall best-performing MLIP on the databases is $\mlip{\text{\rscan-D3}}$, as can be seen in Table~\ref{tab:UAFD_train_MLIP_RMSE_eval}. This also means that the best linear combination of models, given by $\bmw_0$, is close to this MLIP (details in Appendix \ref{sec:si_UAFD_ensemble_weights}), and the performance of the best-performing \petuafd ensemble member is similar to $\mlip{\text{\rscan-D3}}$.
The fact that the MLIPs deviate somewhat from the corresponding DFT calculations on the UAFD training databases is not a major issue per se. Our main focus is to equip MLIPs with error estimation relative to experiment, and therefore the correspondence to the DFT functionals becomes of less importance. The main question is whether the probability distribution of MLIPs constructed from the four databases also provides realistic error estimates for more complicated quantities involving, for example, molecular dynamics simulations. We shall return to this, but first we shall discuss the error estimation on the UAFD training database.

\begin{figure}
\includegraphics[width=\linewidth]{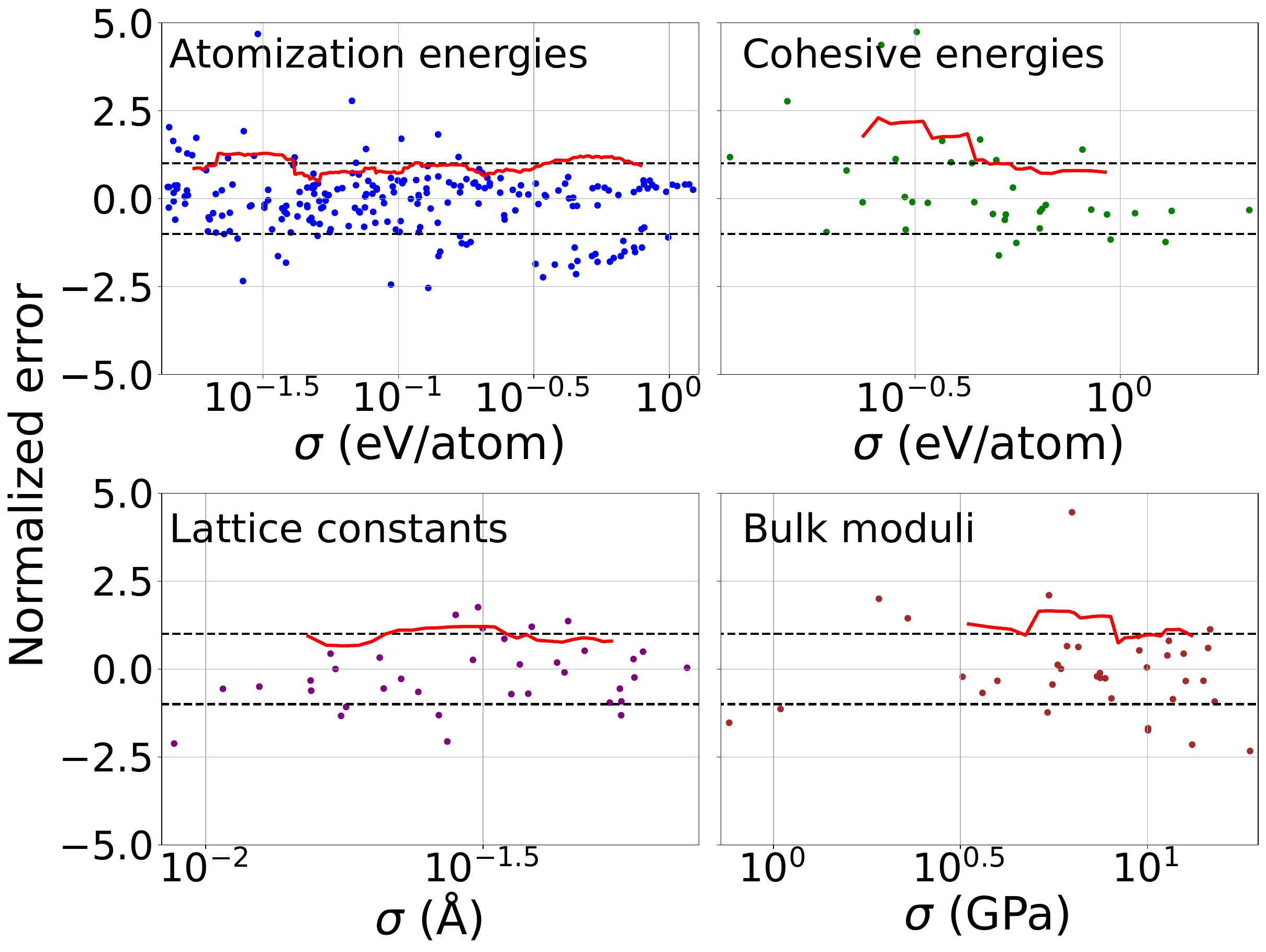}
\caption{Normalized errors vs. uncertainty estimates from cross-validated data. UAFD is 5-fold cross-validated on atomization energies, cohesive energies, lattice constants, and bulk moduli. Running root-mean-square normalized error (RMSNE) (red) is shown in each plot averaged over 30 points in the atomization energies and averaged over 10 points in the rest. Running RMSNE shows that uncertainty estimates are good for atomization energies, lattice constants, and bulk moduli. Cohesive energies have too low uncertainty estimates compared to errors.}
\label{fig:UAFD_uncertainties_4db}
\end{figure}

The predicted uncertainties on the four training databases are illustrated in Fig.~\ref{fig:UAFD_uncertainties_4db}. Here we employ five-fold cross-validation, so the points shown in the figure are collected from the five 20 \% validation sets. The predicted uncertainties (the square root of the variances) are shown on the $x$-axis on a logarithmic scale, while on the $y$-axis we show the normalized error, which is the empirical error (i.e. the difference between the target value and the best model) divided by the predicted uncertainty. 
If we consider a window of predicted uncertainties on the $x$-axis, then -- in an ideal situation -- the normalized errors (the $y$-values) should exhibit a standard normal distribution. To test this, we calculate a running root-mean-square of the normalized error (RMSNE), shown in Fig.~\ref{fig:UAFD_uncertainties_4db} as a red curve. In the ideal case, the red curve should thus be close to one. We see that for the atomization energies, lattice constants, and bulk moduli this is indeed the case, while for the cohesive energies the curve is still reasonable, but the errors tend to be underestimated by about a factor of two in the regime with small uncertainties. It should be noted that the uncertainty prediction for the atomization energies spans two orders of magnitude, demonstrating that the method does not just provide an average uncertainty prediction for each database.
The final PET-UAFD probability distribution is obtained by determining $\bmw_0$ and $\bmK$ using all the data in the four databases without cross-validation. To what extent this distribution is transferable and provides realistic uncertainty estimates for the properties of liquids will be analyzed in the next section.

\subsection{Evaluating \petuafd on the densities and radial distribution functions of liquids}

The cross-validation results in the previous section show that \petuafd provides uncertainty quantification against the experimental ground truth that is comparable to what was obtained in Ref.~\onlinecite{hansenUncertaintyawareElectronicDensityfunctional2025} using direct DFT calculations. 
The true advantage of using an ensemble based on MLIPs becomes clear when quantities are evaluated that require long simulations and large supercells, such as the thermally averaged structural properties of liquids.
As two compelling examples, we consider densities and radial distribution functions of the database of 21 liquids introduced in the Methods Section~\ref{sec:database_substances}.
This is a highly extrapolative exercise, given that the functional distribution is optimized exclusively on the energetic properties of solids and does not even involve all the substances that we consider for these finite-temperature benchmarks.

\begin{table}[b]
    \centering
    \caption{Density prediction errors for different MLIPs.}
    \begin{tabular}{lcc}
        \toprule
        Model & RMSE [g/cm$^3$] & MAE [g/cm$^3$] \\
        \midrule
        $\bm{w}_0$ & 0.44 & 0.30 \\
        $\bm{w}_1$ & 0.44 & 0.31 \\
        $\bm{w}_2$ & 0.71 & 0.42 \\
        $\bm{w}_3$ & 0.44 & 0.30 \\
        $\bm{w}_4$ & 0.35 & 0.26 \\
        \midrule
        $\mlip{\text{PBEsol}}$  & 0.80 & 0.48 \\
        $\mlip{\text{PBE}}$   & 0.56 & 0.41 \\
        $\mlip{\text{PBE+U}}$    & 0.57 & 0.43 \\
        $\mlip{\text{\rscan}}$   & 0.52 & 0.36 \\
        $\mlip{\text{\rscan-D3}}$ & 0.44 & 0.30 \\
        \bottomrule
    \end{tabular}
    \label{tab:density_errors}
\end{table}

\begin{figure}[tbp]
    \centering
    \includegraphics[width=\linewidth]{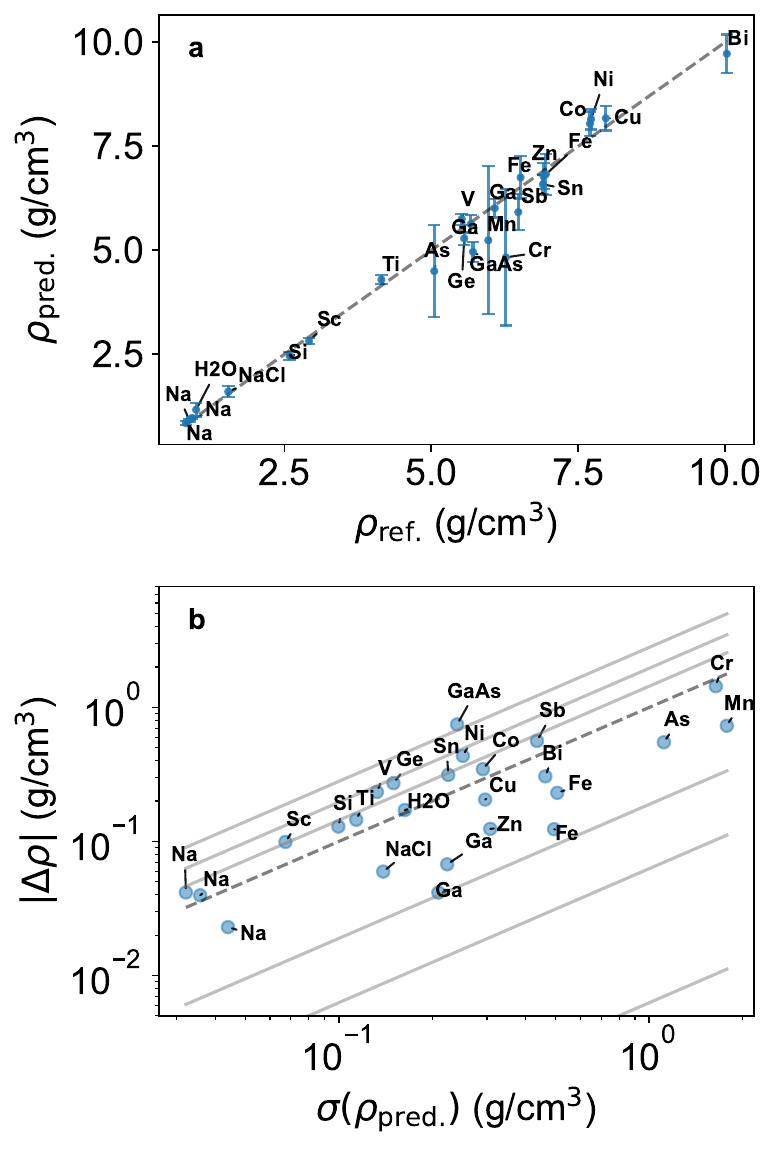}
    \caption{\petuafd predicted densities against experimental reference densities. In subplot a) we plot reference densities against predicted \petuafd densities and indicate the \petuafd uncertainties as error bars. The dashed line indicates the identity line. In subplot b) we plot \petuafd predicted uncertainties $\sigma(\rho_{\text{pred.}})$ against unsigned prediction errors $|\Delta\rho|$. The dashed line indicates the unit line, and gray lines from bottom to top are the 0.5, 5, 15, 85, 95, and 99.5 \% quantile lines of the log folded normal distribution, indicating that between the lowermost and uppermost line 99 \% of the samples should be found. }
    \label{fig:density_error_uq}
\end{figure}

\begin{figure*}[p]
    \centering
    \includegraphics[width=\textwidth]{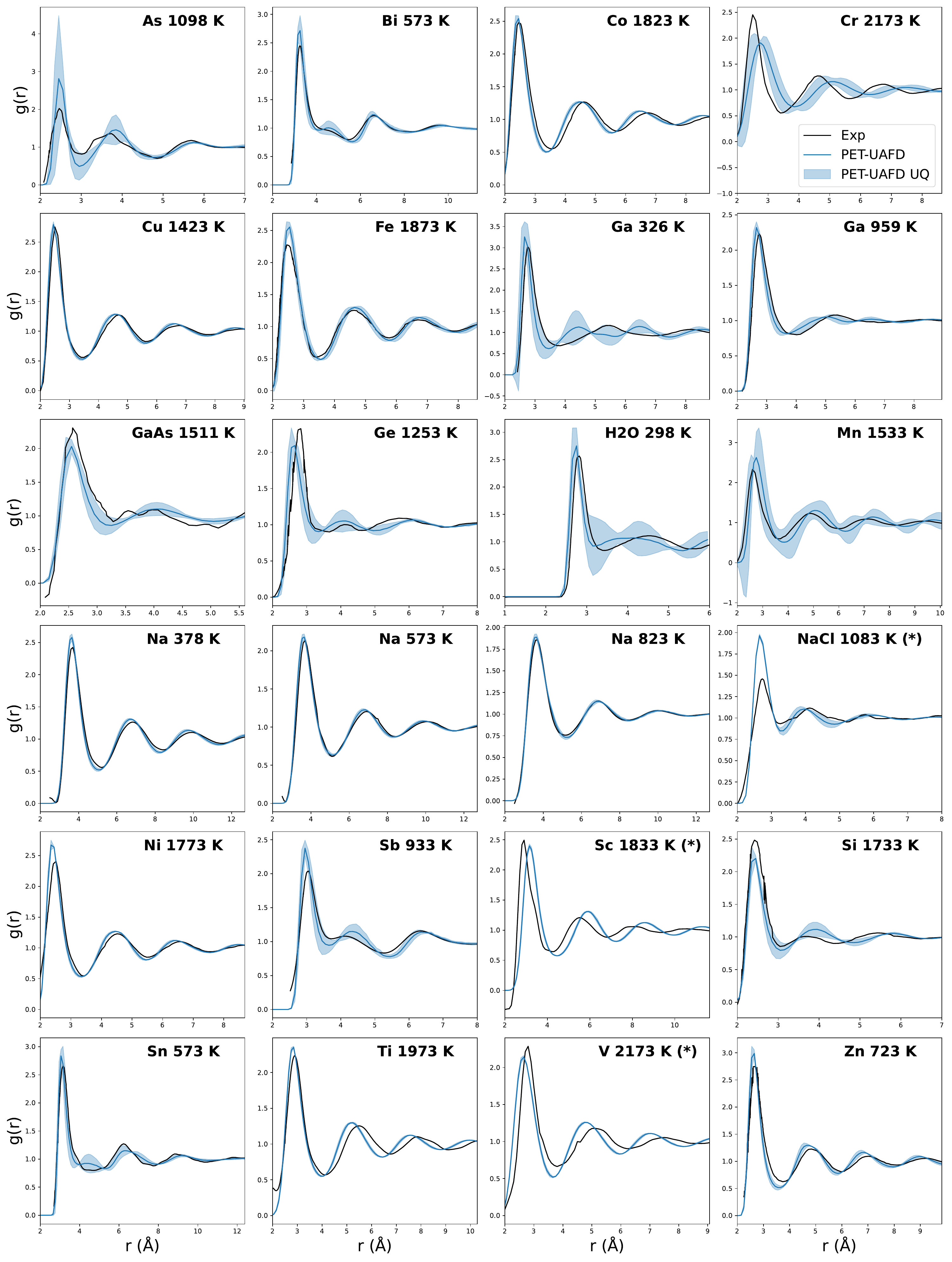}
    \caption{Radial distribution function of selected liquids at ambient pressures ordered alphabetically. \petuafd predicted pair correlation functions are drawn in blue and the \petuafd predicted $\pm$1 standard deviation uncertainties are shaded in blue. Data points that are marked with an asterisk (*) are affected by reported or potentially large measurement uncertainties. More details are given in the main text and the appendix section~\ref{sec:SI_experimental_uncertainty}.}
    \label{fig:RDFs_examples}
\end{figure*}

Studying the densities of the \petuafd simulations allows a quantitative assessment of the \uafd uncertainty estimates and predictions on thermodynamic averages. 
Generally, the simulations of \petuafd are stable across all entries in our database. We also compare the prediction accuracy of the \petuafd with those made by the basis universal MLIPs of \petuafd, which are explicitly trained against PBEsol, PBE, PBE+U, \rscan, and \rscan-D3 (Table~\ref{tab:density_errors}). 
Note that we excluded the liquid arsenic simulation from the analysis, as the PBE, PBE+U and \rscan\ simulations of liquid arsenic diverge due to missing dispersion terms. In general, we find that the density prediction accuracy of the central \petuafd member $\bm{w}_0$ matches the prediction accuracy of the best base functional, the \rscan-D3 member. Interestingly, across our database, the density predictions of the \petuafd ensemble member $\bm{w}_4$ are more accurate than those of $\bm{w}_0$; however, selecting $\bm{w}_4$ for property simulations would not be faithful, since $\bm{w}_0$ was optimized to minimize the prediction errors on the training data and the accuracy for densities might be fortuitous.
Detailed predictions of the individual \petuafd basis and functional MLIPs are listed in the SI section~\ref{sec:si_densities_listing}, Tables~\ref{table:si_uafd_individual_densities} and ~\ref{table:si_xc_individual_densities}.

Across the experimental database, \petuafd prediction errors span more than an order of magnitude in terms of absolute and relative density prediction errors, allowing us to test the quality of the \petuafd uncertainty estimates over scenarios ranging from small prediction errors to substantial prediction errors of 10 \% of the density and more. A good uncertainty estimator should be able to resolve such areas of high- and low-uncertainty predictions, and in line with our observations on static properties, \petuafd can resolve such cases. 
In Figure~\ref{fig:density_error_uq} a) we plot the \uafd predicted densities against the reference values, with error bars indicating the \petuafd predicted standard deviations. In Figure~\ref{fig:density_error_uq} b) we also plot the prediction errors of \petuafd against the uncertainty estimates in a double logarithmic plot. The graphs show that there is a strong general correlation between the predicted uncertainties and the prediction errors across several orders of magnitude. The enveloping lines denote quantiles of the cumulative log folded normal distribution and indicate that for a given predicted uncertainty value, 99 \% of the samples are expected to fall within the outermost enveloping lines.
A more detailed description of how to read this graph can be found in Ref.~\cite{kell-ceri24mlst}.

We already established that the density prediction errors of \petuafd match those of the best xc-uMLIP, namely the \rscan-D3 MLIP. This underlines that the density predictions for which we aim to quantify prediction uncertainty are not artificially made worse by the calibration procedure, and we consider them near the limit of the accuracy
currently achievable with universal
machine-learning potentials trained on
these datasets and architectures. In general, \rscan-D3 MLIP and $\bm{w}_0$ produce density predictions of the highest accuracy, but it is important to stress that the functional choice that gives density estimates in the best agreement with experiment is not always \rscan-D3 MLIP or $\bm{w}_0$. Without prior knowledge of experimental measurements, $\bm{w}_0$ is the best estimator of a liquid's density -- elevated density uncertainty estimates might indicate the need to switch to a higher level of electronic structure theory or carefully calibrate existing DFT functionals on structurally and chemically related compounds.

Although density simulations provide a simple quantitative benchmark for the ability of \petuafd to predict uncertainties in the extrapolative regime, the radial distribution function (RDF) provides a richer characterization of the structure of a liquid. 
In Figure~\ref{fig:RDFs_examples}, we show RDFs computed with \petuafd for all database entries, together with the predicted uncertainty, and compare them to experiments. 
The evaluation of the database reveals cases for which both \petuafd uncertainty estimates and prediction errors are small, such as Na and Ni; moderately elevated, such as Si and Sn; or large, such as Mn, As and Cr. \petuafd uncertainty estimates are able to correctly identify areas of elevated prediction errors compared to experiment. 
Generally, across our database, we observe that the RDF \uafd uncertainties correlate well with the actual prediction errors against experiment. Interestingly, the uncertainty estimates made by \petuafd are not uniform across coordination shells, but in most cases (\emph{i.e.}\ Bi, Ga) correctly identify the areas in the radial distribution functions in which prediction errors are elevated. 
Selected failure cases in which the \petuafd uncertainty estimates are overconfident can be attributed to suspected and reported measurement errors, such as systematic errors in the heights of the first coordination shell in the diffraction measurements \cite{petkovPairDistributionFunctions2012}. 
Specifically for NaCl, the literature states that the position of the first peak is untrustworthy \cite{toveyDFTAccurateInteratomic2020, ohnoXrayDiffractionAnalysis1981}, while for scandium and vanadium, the literature reports that the experimental references must be considered to be affected by non-negligible (but unspecified) measurement errors \cite{delrioFirstPrinciplesDetermination2020}. A more detailed discussion of measurement errors is given in the Appendix Section~\ref{sec:SI_experimental_uncertainty}.

\begin{figure}[b]
\centering\includegraphics[width=1.0\linewidth]{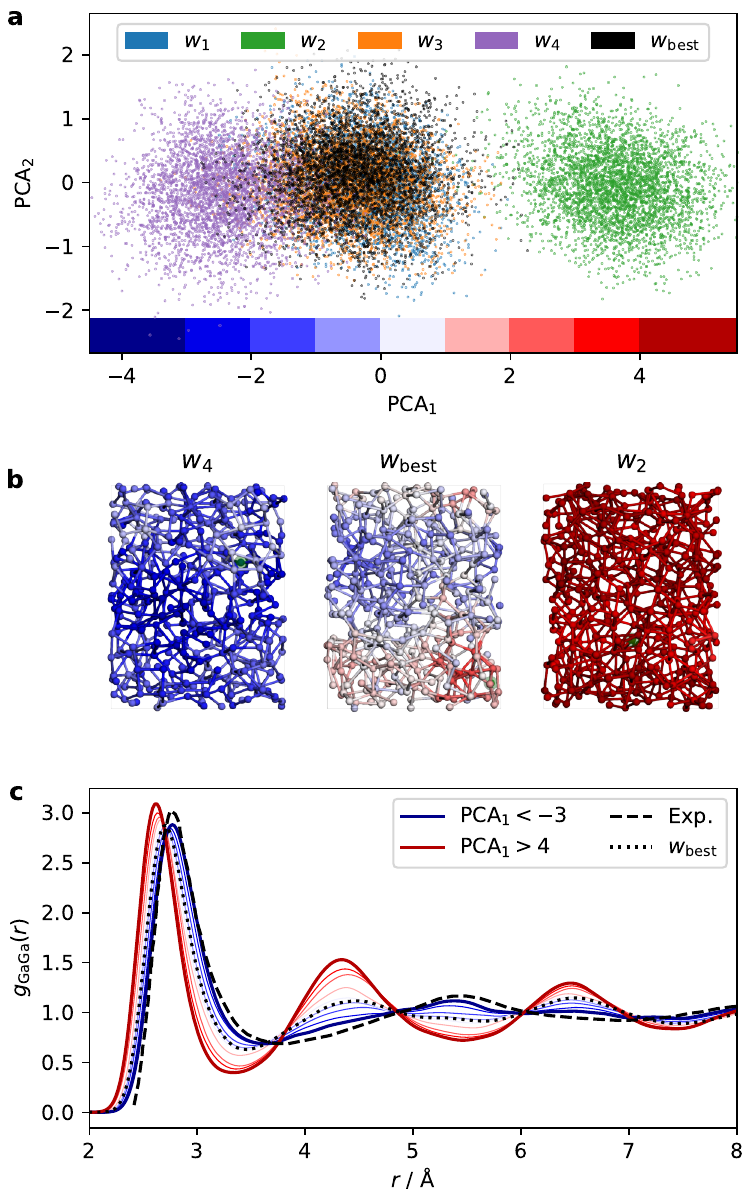}
\caption{Demonstration of the origin of invariant points in the pair correlation functions generated by an MLIP ensemble for liquid Ga at 326~K. 
(a) Principal components of the atom-resolved pair-correlation function $\tilde{g}(r,i)$. The color-coded rectangles correspond to the coloring of atoms in panel b and the average correlation functions in panel c.
(b) Representative snapshots extracted from simulations using the weighted potentials $w_4$, $w_\mathrm{best}$, and $w_2$, color-coded according to the first PCA component.}
\label{fig:isosbestic}
\end{figure}

\subsection{Invariant points in the $g(r)$}

Let us return to the observation that for many of the RDFs in Fig.~\ref{fig:RDFs_examples} the error bars of the pair correlation functions show oscillations and that, intriguingly, the experimental $g(r)$ tends to match simulations at the low-uncertainty points. 
This is, for instance, very evident for \ce{Ga}, \ce{Sb}, \ce{Sn}, and \ce{Cr}, and indicates that the $g(r)$ for all weighted potentials intersect at the same ``invariant'' points. 
In spectroscopy, \emph{isosbestic} points are frequencies for which the intensity of a spectrum is invariant to a change in conditions (e.g. the progress of a chemical reaction). The observation of such points is a strong indication that the signal originates from the superimposition of two spectra that are linearly mixed as the external parameter varies.
To demonstrate that something similar is at play in this case, we compute an atom-centered RDF: 
\begin{equation}
g(r,i) = \sum_j \frac{\Delta(r-r_{ij})}{4\pi r^2 \rho},
\quad
\tilde{g}(r,i)=\frac{\sum_j g(r,j) e^{-\frac{r_{ij}^2}{2\sigma^2}} }{ \sum_j e^{-\frac{r_{ij}^2}{2\sigma^2}} },
\end{equation}
where $\Delta$ is a finite-width Gaussian approximating a Dirac-$\delta$, and spatial convolution is used to average the signal from nearby atoms. 

We compute $\tilde{g}(r,i)$ for a set of representative snapshots from trajectories computed for \ce{Ga} at 326~K with each weighted potential, and perform a principal component analysis (PCA) to classify the atomic environments in a way that emphasizes the variations in local pair correlations. 
The results of this analysis (\Cref{fig:isosbestic}a) show that simulations from different members of the ensemble are clearly clustered along the first PCA axis. The ``best'' potential, as well as the weighted potentials 1 and 3, are in the middle, while the weighted potentials 2 and 4 are at the extremes.
Coloring atoms by the value of the first PCA coordinate (\Cref{fig:isosbestic}b) shows that simulations with intermediate values result in structures that are a mixture of regions with atom-resolved $g(r)$ resembling the two extremes.
Indeed, when we plot averages of the pair correlations within different ranges of the PCA coordinates (\Cref{fig:isosbestic}c), we observe near-perfect crossings, corresponding to the low-uncertainty regions.
We also see that the experimental curve not only passes through the invariant points but is very close to the mean $\tilde{g}(r,i)$ at one of the edges of the distribution. 
This analysis indicates that the liquid contains two types of environments and that the primary failure mode of DFT is to incorrectly estimate the relative population of these two types, explaining the behavior of the predicted uncertainty. 

\subsection{Efficient \uafd uncertainties with \petuafdh}

\begin{figure*}[t]
    \centering
    \includegraphics[width=\textwidth]{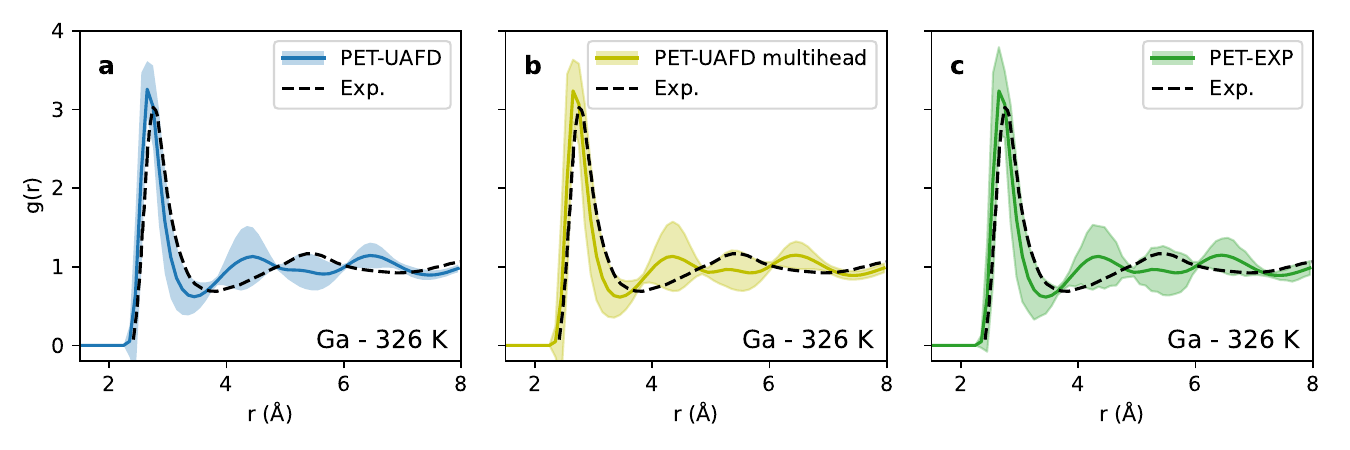}
    \caption{Comparing simulated and experimentally obtained RDFs for liquid gallium at 326~K: a) \petuafd predicted RDF$\pm 1\sigma$ compared with experiment; b) \petuafd multihead model predicted RDF$\pm 1\sigma$ compared to experiment; c) RDF$\pm 1\sigma$ obtained by the full \petuafdh protocol (combining thermodynamic reweighting and \petuafd multihead models) compared to experiment.}
    \label{fig:compare_Ga_pet_uafd_exp}
\end{figure*}

Making predictions of ensemble averages with the \uafd framework incurs higher costs than employing a single universal MLIP, given that each \uafd ensemble member requires the evaluation of $(m+1)$ MLIPs. 
Therefore, when naively applying \petuafd and the \uafd formalism to predict ensemble averages, the evaluation scales quadratically with the number of basis functionals, as a separate molecular dynamics simulation has to be performed for each of the ensemble members. Although the number of basis functionals in \petuafd is relatively small, and the MLIPs are far less computationally demanding than even the non-self-consistent evaluation of density functional theory energetics, the additional cost is still significant and needs to be carefully traded off against longer sampling durations and/or larger simulation cells.
We already discussed in Section~\ref{sec:methods} how this overhead can be eliminated almost entirely using a shallow ensemble and statistical reweighting. 
To demonstrate how this works in practice in the case of the radial distribution functions of liquid gallium at 326~K, we proceed in steps.
First, we show that the multi-head model reproduces the error estimates of the full \petuafd ensemble faithfully when performing simulations separately using the outputs of each head (Fig.~\ref{fig:compare_Ga_pet_uafd_exp}b). 
Consistent with the small changes in the accuracy of the MLIP predictions for the multi-head fit, we find that predicted radial distribution functions and uncertainty estimates are in good agreement with the explicit evaluation of separate models in \petuafd, at a greatly reduced cost.

We then deploy the \petuafdh protocol in full by performing a single trajectory using the head corresponding to $\bm{w}_0$ and then generating the RDF corresponding to the other \uafd ensemble members using the statistically robust approximate reweighting in Eq.~(\ref{eq:reweight-cea}) (Figure~\ref{fig:compare_Ga_pet_uafd_exp}c), again for the case of liquid Ga. 
We find that the uncertainty estimates from independent simulations
and from reweighting are in almost perfect agreement with the \petuafd results, confirming that thermodynamic reweighting is a viable alternative for uncertainty-quantification in place of running full and independent molecular dynamics simulations.

\section{Conclusion}

In conclusion, we have demonstrated how an ensemble of universal machine-learning potentials trained on different DFT functionals can be calibrated against experimental data. 
The ensemble yields meaningful error estimates against experimental measurements. 
Calibrating the ensemble of universal MLIPs did not degrade the prediction accuracy compared to the most accurate ensemble member MLIP. 
The superior evaluation speed of universal MLIPs allowed us to compute uncertainty estimates of uncertainty estimates for liquid densities and radial
distribution functions. 
These quantities were not part of the \uafd calibration
training set, demonstrating transferability of the calibration protocol to previously unseen experimental quantities. This is crucial because thermodynamic averages are typically the ultimate targets of atomic-scale simulations, and because experimental measurements of them is even sparser and subject to larger experimental uncertainty than the simple structural and energetic quantities we use for calibration.

The explicit evaluation of an ensemble of predictions still adds a substantial overhead over a conventional MLIP simulation, and such overhead grows steeply when using a larger, more diverse set of functionals -- which is likely to be highly beneficial in terms of the accuracy of the error estimator.
We demonstrated that combining UAFD with the DPOSE uncertainty-quantification method reduces the computational overhead of evaluating the errors in derived quantities to a negligible overhead compared with a single MLIP simulation with similar architecture and hyperparameters.
There is still much room for improvement, from using more diverse and consistently curated MLIP training datasets to including more MLIPs in the UAFD basis, more experimental targets in the calibration and accounting for experimental errors, and ultimately assessing the transferability to other, even more challenging targets such as phase-transition temperatures or
time-dependent materials properties. 
Still, it is hard to overstate the utility of a universal MLIP that is capable of estimating the accuracy of simulations against experiments rather than against electronic-structure calculations of unknown fidelity, and doing so at virtually no extra cost.

\section{Data and code availability}
Model checkpoints and calibration data will be made available from a Materials Cloud deposit at XXX.
Additionally, we will make \petuafd and \petuafdh models available via the UPET PyPI package at \url{github.com/lab-cosmo/upet}.

\begin{acknowledgments}
We thank Filippo Bigi for assisting with the metatrain fine-tuning infrastructure and making early OMAT checkpoints available to us.
M.K. and M.C. acknowledge financial support by the Swiss National Science Foundation (Project 200020 214879). We acknowledge support from the Novo Nordisk Foundation Data Science Research Infrastructure 2022 Grant: A high-performance computing infrastructure for data-driven research on sustainable energy materials, Grant no. NNF22OC0078009.
\end{acknowledgments}

\clearpage

\onecolumngrid
\part*{Supplementary Information}

\setcounter{secnumdepth}{3}           %
\renewcommand{\thesection}{S\arabic{section}}
\setcounter{section}{0}

\renewcommand{\thefigure}{S\arabic{figure}}
\setcounter{figure}{0}

\renewcommand{\thetable}{S\arabic{table}}
\setcounter{table}{0}

\section{MLIP model architecture and training details}
\label{sec:SI_training_details}
In this work, an ensemble of universal machine-learning interatomic potentials (uMLIPs) is generated by training point-edge transformer models (PET) \cite{pozd-ceri23nips} on universal databases. Following previous work by Bigi et al. \cite{bigiPushingLimitsUnconstrained2026}, we apply a fine-tuning strategy in which model weights are not randomly initialized, but rather initialized from a set of weights obtained by training on the OMAT24 database. In the training procedure, all weights are inherited, and an additional composition model is fitted to normalize the target DFT labels and make them compatible for fine-tuning. In addition to substantially reducing
training costs (all training runs finished within 48 GPU-hours), this approach provides
two benefits: First, all models benefit from transferring knowledge from the vast corpus of OMAT training structures. Second, in a multihead setting in which the PBE+U backbone remains frozen and only new heads are attached and fine-tuned, the PBE+U head which is inherited from the originally trained model, remains unchanged.
Note that unlike in the original publication by Bigi et al. \cite{bigiPushingLimitsUnconstrained2026}, these models do not employ an adaptive cutoff strategy, and use a fixed cutoff of 4.5 \AA\, across the entire domain of applicability of the MLIP. During head-only fine-tuning, an additional read-out head in the PET model is created and the weights of the original OMAT-trained heads are inherited. During training, only the parameters of the new read-out head are optimized and the backbone weights remain fixed. In Table~\ref{tab:hypers_train} we list training hyperparameters and details of the exact PET architecture used. All training runs were performed with a development version of \texttt{metatrain} corresponding to commit \texttt{5e70194a9a5bbbc1779d5b18b40d8639431a3443}. Features for training multihead potentials are now also available in the main branch of the \texttt{metatrain} library.

\begin{table}[ht]
    \centering
    \caption{Training and fine-tuning hyperparameters as defined in the \texttt{metatrain} library. \cite{bigi+26jcp}}
    \label{tab:hypers_train}
    \begin{tabular}{lc}
    \toprule
    Parameter & Value \\
    \midrule
    Total number of parameters & 195M \\
    Number of fine-tuned parameters (per head) & 7.4 M (3.8 \%)\\
    \midrule
    \texttt{cutoff} & 4.5 \AA \\
    \texttt{d\_pet} & 512 \\
    \texttt{d\_head} & 512 \\
    \texttt{num\_gnn\_layers} & 3 \\
    \texttt{num\_attention\_layers} & 2 \\
    \midrule
    \texttt{batch\_size} & 16 \\
    \texttt{learning\_rate} & 3e-5 \\
    \texttt{grad\_clip\_norm} & 1.0 \\
    \midrule
    \texttt{loss:\,type:} & \texttt{huber} \\
    \texttt{energy:\,delta:} & 0.015 \\
    \texttt{gradients:\,positions:\,delta:} & 0.040 \\
    \texttt{gradients:\,strain:\,delta:} & 0.030 \\
    \texttt{gradients:\,positions:\,weight:} & 1.0 \\
    \texttt{gradients:\,strain:\,weight:} & 1.0 \\
    \texttt{energy:\,weight:} & 1.0 \\
    \midrule
    \texttt{num\_epochs:} & 50\,(MATPES), 100\,(MAD) \\
    \bottomrule
    \end{tabular}
\end{table}

\section{Evaluation of xc-Ensemble-uMLIPs}
\label{sec:training_databases}
We evaluate the baseline universal MLIPs that form \petuafd (independent models) and \petuafdh (backbone shared) on hold-out test sets, to assess the direct prediction accuracy of both ensembles on hold-out reference DFT energy and force labels. In Table~\ref{tab:full_test_eval} we list energy and force prediction errors of the respective uMLIP against hold-out data of the respective DFT functional against which they are
trained. In Table~\ref{tab:head_test_eval} we list test set errors of both committees.

\begin{table}[ht]
    \centering
    \caption{Evaluation metrics of the \petuafd basis universal MLIPs on hold-out test sets}
    \label{tab:full_test_eval}
    \resizebox{0.45\textwidth}{!}
    {\begin{tabular}{lcc}
        \toprule
        Models & E MAE [meV/atom] & F MAE [meV/\AA]\\
        \midrule
        $\mlip{\text{PBE}}$   & 16.0 & 54.8 \\
        $\mlip{\text{PBE+U}}$ & 12.1 & 53.9 \\
        $\mlip{\text{PBEsol}}$ & 4.6 & 28.5 \\
        $\mlip{\text{\rscan}}$ & 14.2 & 70.8 \\
        $\mlip{\text{\rscan-D3}}$  & 14.9 & 72.7 \\
        \bottomrule
    \end{tabular}}
\end{table}

\begin{table}[ht]
    \centering
    \caption{Evaluation metrics of the \petuafdh basis universal MLIPs on hold-out test sets. The asterisk denotes the PBE+U model, whose head predictions will match exactly those of the full model, given that all head-only fine-tuned models share the same PBE+U backbone. }
    \label{tab:head_test_eval}
    \resizebox{0.45\textwidth}{!}
    {\begin{tabular}{lcc}
        \toprule
        Models & E MAE [meV/atom] & F MAE [meV/\AA]\\
        \midrule
        $\mlip{\text{PBE}}$      & 20.4 & 62.2 \\
        $\mlip{\text{PBE+U}}$ & 12.1 & 53.9 \\
        $\mlip{\text{PBEsol}}$  & 8.7 & 44.3 \\
        $\mlip{\text{\rscan}}$  & 19.2 & 78.4 \\
        $\mlip{\text{\rscan-D3}}$   & 20.1 & 80.7 \\
        \bottomrule
    \end{tabular}}
\end{table}

\section{MLIP training sets}
Throughout this work we make use of the MATPES databases (PBE and \rscan) \cite{kaplanFoundationalPotentialEnergy2025a}, the OMAT database \cite{barroso-luqueOpenMaterials20242024} (PBE+U), as well as the MAD database (v1.0, PBEsol) \cite{mazi+25ncomm, mazi+25sd} to train universal MLIPs on different reference DFT computations. In Table~\ref{tab:summary_training_data} we list an overview of the training, validation and test set sizes as well as the \petuafd uMLIP prediction accuracy on the test set.

\begin{table}[ht]
    \centering
    \caption{Evaluation metrics of full-PET-uMLIP on hold-out test set, for the OMAT24 checkpoint the evaluation MAEs on the official validation split are reported.}
    \label{tab:summary_training_data}
    \resizebox{0.75\textwidth}{!}
    {\begin{tabular}{l|c|ccc|cc}
        \toprule
        Functional & Database & n$_{\text{train}}$ & n$_{\text{val}}$ & n$_{\text{test}}$ & E MAE [meV/atom] & F MAE [meV/\AA]\\
        \midrule
        PBE     & MATPES & 391242 & 21735 & 21735 & 16.0 & 54.8 \\
        PBE+U$^{\star}$ & OMat24 & $\sim$100M & $\sim$5M & - &12.1 & 53.9 \\
        PBEsol & MAD1.0 & 60019 & 7514 & 7485 & 4.6 & 28.5 \\
        $\rscan$ & MATPES & 349109 & 19394 & 19394 & 14.2 & 70.8 \\
        $\rscan$-D3  & MATPES & 349109 & 19394 & 19394 & 14.9 & 72.7 \\
        \bottomrule
    \end{tabular}}
\end{table}

\section{UAFD Outlier removal}
\label{sec:si_uafd_outlier_removal}
The solids K, Rb, Sr, Cs and Ba are removed from the solids dataset due to the MLIPs' cutoff radius being too small for these systems. The molecules Li$_2$ and F$_2$ are outliers in the training data and are therefore removed. Li$_2$ has too low a variance over the basis models compared with the best prediction and target value. The PBEsol MLIP prediction for the F$_2$ atomization energy is an outlier and the data point is removed. The solids C and V are removed from the training set because they have low predictive variancev in the bulk moduli while exhibiting large prediction errors against experiment, resulting in an outlier data point. The same applies to solid Ca, except for
the cohesive energy.

\section{UAFD ensemble weights }
\label{sec:si_UAFD_ensemble_weights}
In Table~\ref{tab:exact_UAFD_ensemble_weights} we list the weights of the \petuafd and \petuafdh ensembles.

\begin{table}[h!]
    \centering
    \caption{Weights used to construct the explicit \petuafd and \petuafd multihead models.}
    \begin{tabular}{c|c|c|c|c|c}
        \toprule
        Model & $\mlip{\text{PBE}}$ & $\mlip{\text{PBEsol}}$ & $\mlip{\text{PBE+U}}$ & $\mlip{\text{\rscan}}$ & $\mlip{\text{\rscan-D3}}$ \\
        \midrule
        $\bm{w}_0$ &  0.117 & -0.077 & -0.006 & -0.044 & 1.010  \\
        $\bm{w}_1$ & -0.043 &  0.109 & -0.063 & -0.659 & 1.656  \\
        $\bm{w}_2$ & -0.598 & -0.079 &  0.247 &  0.574 & 0.856 \\
        $\bm{w}_3$ &  0.119 & -0.086 &  0.003 & -0.051 & 1.014 \\
        $\bm{w}_4$ &  0.165 &  0.061 &  0.119 & -0.338 & 0.993 \\
        \bottomrule
    \end{tabular}
    \label{tab:exact_UAFD_ensemble_weights}
\end{table}

\section{Detailed $\mlip{i}$ evaluation results}

\begin{figure}[h!]
    \centering
    \includegraphics[width=0.37\linewidth]{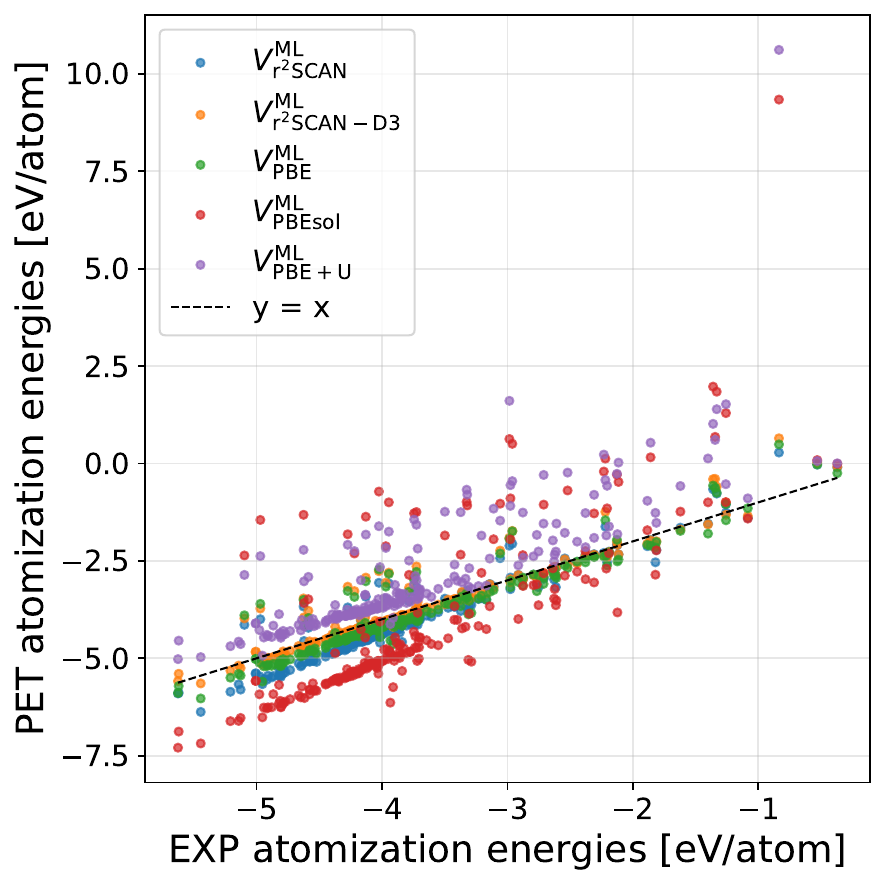}
    \includegraphics[width=0.37\linewidth]{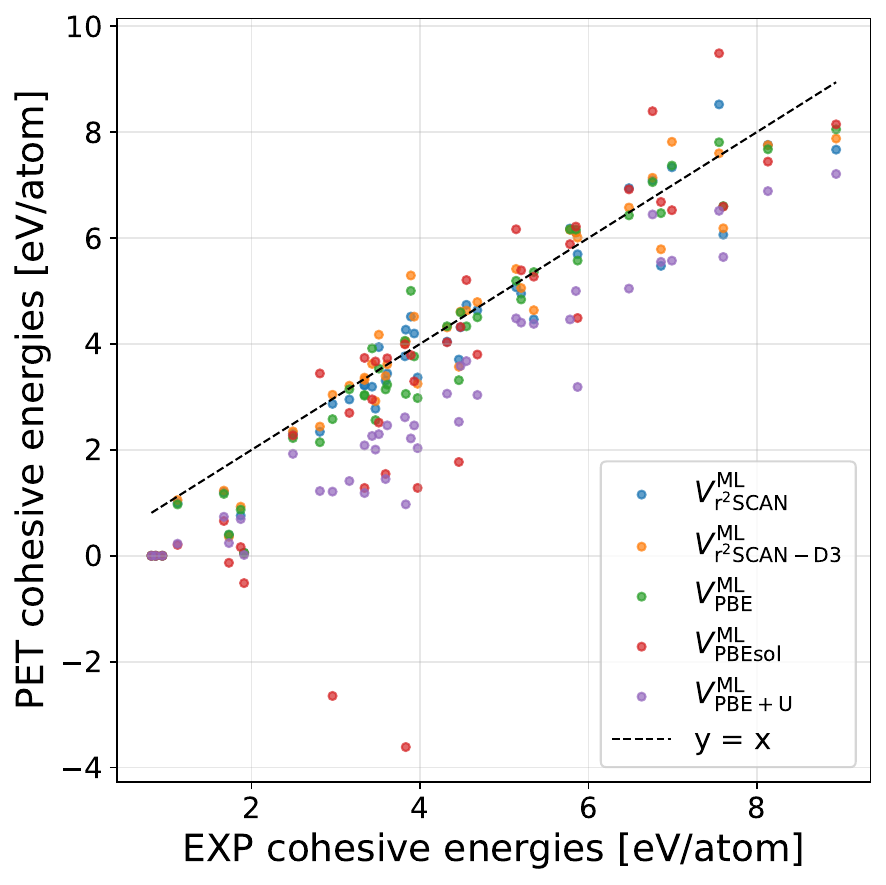}\\
    \includegraphics[width=0.37\linewidth]{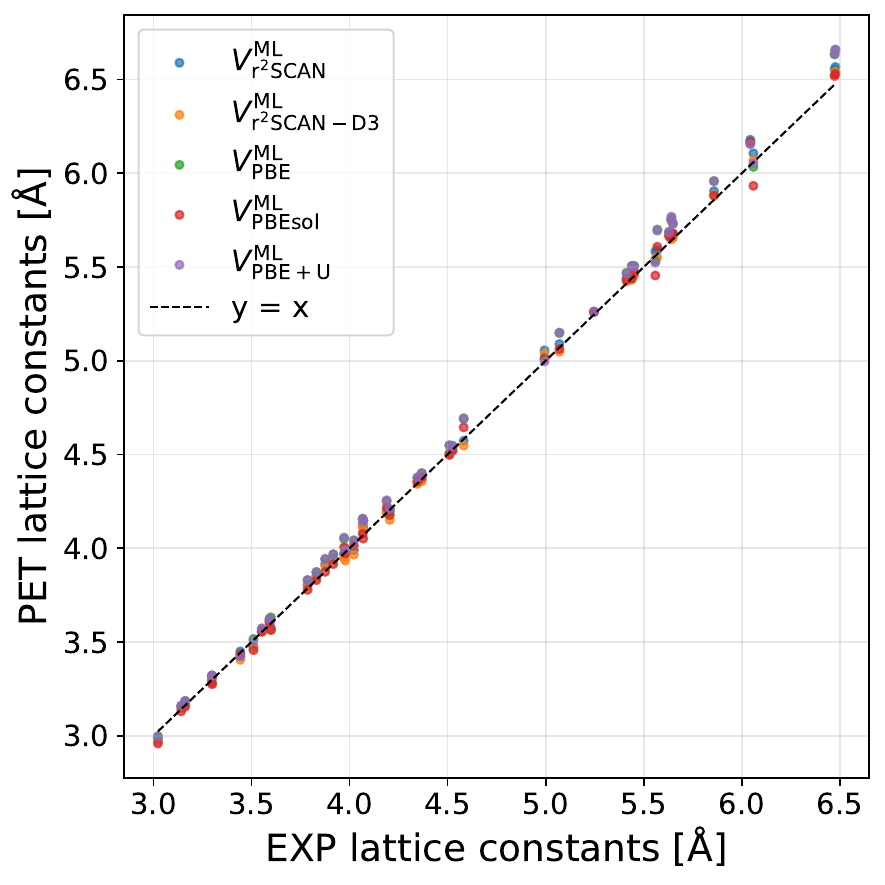}
    \includegraphics[width=0.37\linewidth]{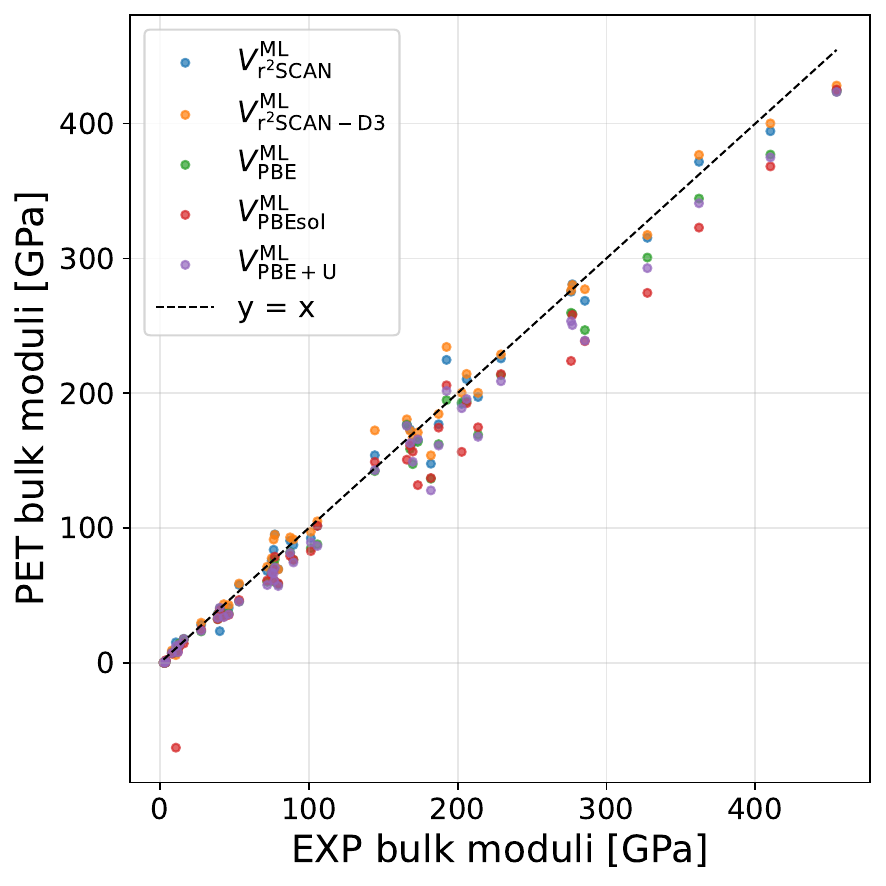}
    \caption{Parity plots comparing MLIP $\mlip{i}$ predictions with experimental values. The comparison is shown for atomization energies, cohesive energies, lattice constants and bulk moduli.}
    \label{fig:UAFD_basis_pred}
\end{figure}

In Figure~\ref{fig:UAFD_basis_pred} we show parity plots of the individual \uafd basis MLIPs  $\mlip{i}$ against experimental reference values introduced in Ref.~\citenum{hansenUncertaintyawareElectronicDensityfunctional2025}.

\newpage
\section{Experimental reference values} \label{sec:experimental_references}
In Table~\ref{tab:si_list_experimental_references} we list the experimental sources of the experimental reference database that we curated.

\begin{table}[h!]
    \centering
    \caption{Experimental references in the database}
\label{tab:si_list_experimental_references}
\begin{tabular}{l|c|c|c|c}
\toprule
Species & T [$\mathrm{K}$] & $\rho_{\mathrm{ref}}$ [$\mathrm{g\,cm^{-3}}$] & Experimental source RDF & Experimental source $\rho$\\
\midrule
H2O  & 298  & 0.997 & Soper \cite{soperRadialDistributionFunctions2000c} & CRC-handbook \cite{CRCDensityWater2005} \\
As   & 1098 & 5.05  & Bellissent \cite{bellissentStructureLiquidPeierls1987} & Klemm \cite{klemmFurtherContributionsKnowledge1963} \\
Ga   & 326  & 6.08  & Bellissent-Funel \cite{bellissent-funelStructureFactorEffective1989} &  Bellissent-Funel \cite{bellissent-funelStructureFactorEffective1989}\\
Ga   & 959  & 5.674 & Bellissent-Funel \cite{bellissent-funelStructureFactorEffective1989} & Bellissent-Funel \cite{bellissent-funelStructureFactorEffective1989}\\
GaAs & 1511 & 5.71  & Bergman \cite{bergmanATOMICSTRUCTURELIQUID1985} & Glazov \cite{Glazov1969LiquidSemiconductors} \\
Na   & 378  & 0.928 & Waseda \cite{waseda1980structure} & Waseda \cite{waseda1980structure} \\
Na   & 573  & 0.881 & Waseda \cite{waseda1980structure} & Waseda \cite{waseda1980structure} \\
Na   & 823  & 0.823 & Waseda \cite{waseda1980structure} & Waseda \cite{waseda1980structure} \\
Zn   & 723  & 6.91  & Lou \cite{louNegativeExpansionsInteratomic2013} & Waseda \cite{waseda1980structure} \\
Cu   & 1423 & 7.97  & Waseda \cite{waseda1980structure} & Waseda \cite{waseda1980structure} \\
Ni   & 1773 & 7.72  & Schmitz-Pranghe \cite{schmitz-prangheNotizenRontgenbeugungsuntersuchungenEisen1970} & Waseda \cite{waseda1980structure}\\
Co   & 1823 & 7.7   & Schmitz-Pranghe \cite{schmitz-prangheNotizenRontgenbeugungsuntersuchungenEisen1970} & Waseda \cite{waseda1980structure}\\
Fe   & 1873 & 6.95  & Mendelev, Il'inskii \cite{mendelevDevelopmentNewInteratomic2003, ilinskiiStructureLiquidFe2002} & Waseda \cite{waseda1980structure}\\
Fe   & 2023 & 6.52  & - & Waseda \cite{waseda1980structure}\\
Mn   & 1533 & 5.97  & Waseda \cite{waseda1980structure} & Waseda \cite{waseda1980structure}\\
Cr   & 2173 & 6.27  & Waseda \cite{waseda1980structure} & Waseda \cite{waseda1980structure}\\
V    & 2173 & 5.517  & Waseda \cite{waseda1980structure} & Ntonti \cite{ntontiReferenceCorrelationsDensity2024}\\
Ti   & 1973 & 4.15  & Holland-Moritz \cite{holland-moritzShortrangeOrderStable2007} & Waseda \cite{waseda1980structure}\\
Sc   & 1833 & 2.92  & Waseda \cite{waseda1980structure} & Waseda \cite{waseda1980structure} \\
Bi   & 573  & 10.03 & Mayo, Greenberg \cite{mayoShortRangeOrder2013, greenbergEvidenceTemperaturedrivenStructural2009} & Waseda \cite{waseda1980structure} \\
Sn   & 573  & 6.9   & Itami \cite{itamiStructureLiquidSn2003} & Waseda \cite{waseda1980structure} \\
Ge   & 1253 & 5.56  & Waseda \cite{waseda1980structure} & Waseda \cite{waseda1980structure} \\
Si   & 1733 & 2.59  & Waseda \cite{waseda1980structure} & Waseda \cite{waseda1980structure} \\
NaCl & 1083 & 1.545 & Ohno \cite{ohnoXrayDiffractionAnalysis1981} & Ohno \cite{ohnoXrayDiffractionAnalysis1981} \\
Sb   & 933  & 6.48  & Mayo, Greenberg \cite{mayoShortRangeOrder2013, greenbergRelationMicroscopicStructure2010} & Waseda \cite{waseda1980structure}\\
\bottomrule
\end{tabular}
\end{table}

\section{\petuafd density predictions}
\label{sec:si_densities_listing}

In Table~\ref{table:si_uafd_individual_densities} we list the individual density predictions of \petuafd committee members and the \petuafd uncertainty estimates.

\begin{table}[p]
    \centering
    \caption{PET-UAFD predicted densities}
\resizebox{1.\textwidth}{!}{
\begin{tabular}{l|c|c|cccccc}
\toprule
Species & T [$\mathrm{K}$] & $\rho_{\mathrm{ref}}$ [$\mathrm{g\,cm^{-3}}$] & $\rho_{\mathrm{pred}}^{\bm{w}_0}$ [$\mathrm{g\,cm^{-3}}$] & $\rho_{\mathrm{pred}}^{\bm{w}_1}$ [$\mathrm{g\,cm^{-3}}$] & $\rho_{\mathrm{pred}}^{\bm{w}_2}$ [$\mathrm{g\,cm^{-3}}$] & $\rho_{\mathrm{pred}}^{\bm{w}_3}$ [$\mathrm{g\,cm^{-3}}$] & $\rho_{\mathrm{pred}}^{\bm{w}_4}$ [$\mathrm{g\,cm^{-3}}$] & $\sigma(\rho_{\mathrm{pred}})$ [$\mathrm{g\,cm^{-3}}$] \\
\midrule
H2O & 298 & 0.997 & 1.168 & 1.248 & 1.282 & 1.175 & 1.083 & 0.163 \\
As & 1098 & 5.05 & 4.501 & 5.037 & 3.562 & 4.506 & 4.768 & 1.114 \\
Ga & 326 & 6.08 & 6.012 & 6.212 & 5.917 & 6.014 & 6.043 & 0.223 \\
Ga & 959 & 5.674 & 5.632 & 5.821 & 5.722 & 5.628 & 5.616 & 0.209 \\
GaAs & 1511 & 5.71 & 4.964 & 5.201 & 4.999 & 4.957 & 4.986 & 0.240 \\
Na & 378 & 0.928 & 0.970 & 1.002 & 0.965 & 0.970 & 0.970 & 0.032 \\
Na & 573 & 0.881 & 0.921 & 0.956 & 0.917 & 0.921 & 0.918 & 0.036 \\
Na & 823 & 0.823 & 0.846 & 0.889 & 0.845 & 0.846 & 0.837 & 0.044 \\
Zn & 723 & 6.91 & 6.786 & 7.041 & 6.949 & 6.778 & 6.733 & 0.307 \\
Cu & 1423 & 7.97 & 8.175 & 8.386 & 8.379 & 8.171 & 8.129 & 0.296 \\
Ni & 1773 & 7.72 & 8.155 & 8.332 & 8.332 & 8.146 & 8.142 & 0.251 \\
Co & 1823 & 7.7 & 8.046 & 8.312 & 8.076 & 8.038 & 8.160 & 0.291 \\
Fe & 1873 & 6.95 & 6.827 & 7.204 & 6.768 & 6.810 & 7.139 & 0.494 \\
Fe & 2023 & 6.52 & 6.751 & 7.137 & 6.695 & 6.735 & 7.071 & 0.505 \\
Mn & 1533 & 5.97 & 5.240 & 5.384 & 4.522 & 5.236 & 6.860 & 1.778 \\
Cr & 2173 & 6.27 & 4.831 & 4.905 & 3.648 & 4.832 & 5.962 & 1.638 \\
V & 2173 & 5.517 & 5.751 & 5.881 & 5.753 & 5.748 & 5.776 & 0.133 \\
Ti & 1973 & 4.15 & 4.295 & 4.399 & 4.269 & 4.293 & 4.334 & 0.114 \\
Sc & 1833 & 2.92 & 2.821 & 2.885 & 2.812 & 2.820 & 2.838 & 0.067 \\
Bi & 573 & 10.03 & 9.724 & 10.138 & 9.563 & 9.715 & 9.851 & 0.463 \\
Sn & 573 & 6.9 & 6.586 & 6.798 & 6.538 & 6.584 & 6.645 & 0.225 \\
Ge & 1253 & 5.56 & 5.288 & 5.435 & 5.293 & 5.290 & 5.317 & 0.150 \\
Si & 1733 & 2.59 & 2.461 & 2.540 & 2.417 & 2.459 & 2.503 & 0.100 \\
NaCl & 1083 & 1.545 & 1.605 & 1.709 & 1.684 & 1.601 & 1.560 & 0.139 \\
Sb & 933 & 6.48 & 5.918 & 6.293 & 5.755 & 5.899 & 6.065 & 0.435 \\
\bottomrule
\end{tabular}
}
\label{table:si_uafd_individual_densities}
\end{table}

In Table~\ref{table:si_xc_individual_densities} we list predictions of the \petuafd basis uMLIPs $\mlip{i}$. Note that the $\mlip{\text{PBE}}$, $\mlip{\rscan}$ and $\mlip{\text{PBE+U}}$ MD simulations diverge, presumably due to missing dispersion terms.

\begin{table}[p]
    \centering
    \caption{\petuafd basis MLIP predicted densities and reference densities. }
\resizebox{1.\textwidth}{!}{
\begin{tabular}{l|c|c|ccccc}
\toprule
Species & Temperature [$\mathrm{K}$] & $\rho_{\mathrm{ref}}$ [$\mathrm{g\,cm^{-3}}$] & $\rho_{\mathrm{pred}}^{\mlip{\rscan}}$ [$\mathrm{g\,cm^{-3}}$] & $\rho_{\mathrm{pred}}^{\mlip{\rscan\text{-D3}}}$ [$\mathrm{g\,cm^{-3}}$] & $\rho_{\mathrm{pred}}^{\mlip{\text{PBE}}}$ [$\mathrm{g\,cm^{-3}}$] & $\rho_{\mathrm{pred}}^{\mlip{\text{PBEsol}}}$ [$\mathrm{g\,cm^{-3}}$] & $\rho_{\mathrm{pred}}^{\mlip{\text{PBE+U}}}$ [$\mathrm{g\,cm^{-3}}$] \\
\midrule
As & 1098 & 5.05 & 0.752 & 4.568 & 0.316 & 5.185 & 0.295 \\
Ga & 326 & 6.08 & 5.887 & 6.049 & 5.744 & 6.165 & 5.720 \\
Ga & 959 & 5.674 & 5.534 & 5.676 & 5.265 & 5.740 & 5.258 \\
Na & 378 & 0.928 & 0.930 & 0.971 & 0.920 & 0.942 & 0.912 \\
Na & 573 & 0.881 & 0.874 & 0.922 & 0.856 & 0.864 & 0.806 \\
Na & 823 & 0.823 & 0.792 & 0.847 & 0.748 & 0.478 & 0.515 \\
GaAs & 1511 & 5.71 & 4.773 & 5.010 & 4.550 & 5.205 & 4.535 \\
Fe & 1873 & 6.95 & 6.867 & 6.961 & 7.084 & 8.503 & 7.142 \\
Fe & 2023 & 6.52 & 6.801 & 6.892 & 7.017 & 8.432 & 7.080 \\
NaCl & 1083 & 1.545 & 1.501 & 1.624 & 1.273 & 1.445 & 1.257 \\
Cu & 1423 & 7.97 & 8.075 & 8.234 & 7.629 & 8.197 & 7.700 \\
Mn & 1533 & 5.97 & 5.342 & 5.298 & 7.594 & 8.331 & 7.602 \\
V & 2173 & 5.517 & 5.668 & 5.779 & 5.642 & 5.912 & 5.638 \\
Cr & 2173 & 6.27 & 4.770 & 4.820 & 6.665 & 6.954 & 6.667 \\
Bi & 573 & 10.03 & 9.253 & 9.779 & 9.423 & 10.248 & 9.372 \\
Ti & 1973 & 4.15 & 4.216 & 4.311 & 4.243 & 4.461 & 4.257 \\
Sn & 573 & 6.9 & 6.415 & 6.623 & 6.393 & 6.890 & 6.391 \\
Co & 1823 & 7.7 & 8.023 & 8.131 & 7.929 & 8.894 & 7.933 \\
Zn & 723 & 6.91 & 6.655 & 6.848 & 6.191 & 6.868 & 6.152 \\
Ni & 1773 & 7.72 & 8.133 & 8.218 & 7.805 & 8.393 & 7.863 \\
Sc & 1833 & 2.92 & 2.762 & 2.829 & 2.763 & 2.901 & 2.776 \\
Si & 1733 & 2.59 & 2.415 & 2.476 & 2.458 & 2.637 & 2.464 \\
Ge & 1253 & 5.56 & 5.234 & 5.327 & 5.138 & 5.572 & 5.114 \\
Sb & 933 & 6.48 & 5.474 & 5.972 & 5.693 & 6.452 & 5.666 \\
\bottomrule
\end{tabular}
}
\label{table:si_xc_individual_densities}
\end{table}

\newpage
\section{Comment on experimental uncertainties}
\label{sec:SI_experimental_uncertainty}
\begin{figure}[t]
    \centering
    \includegraphics[width=0.75\linewidth]{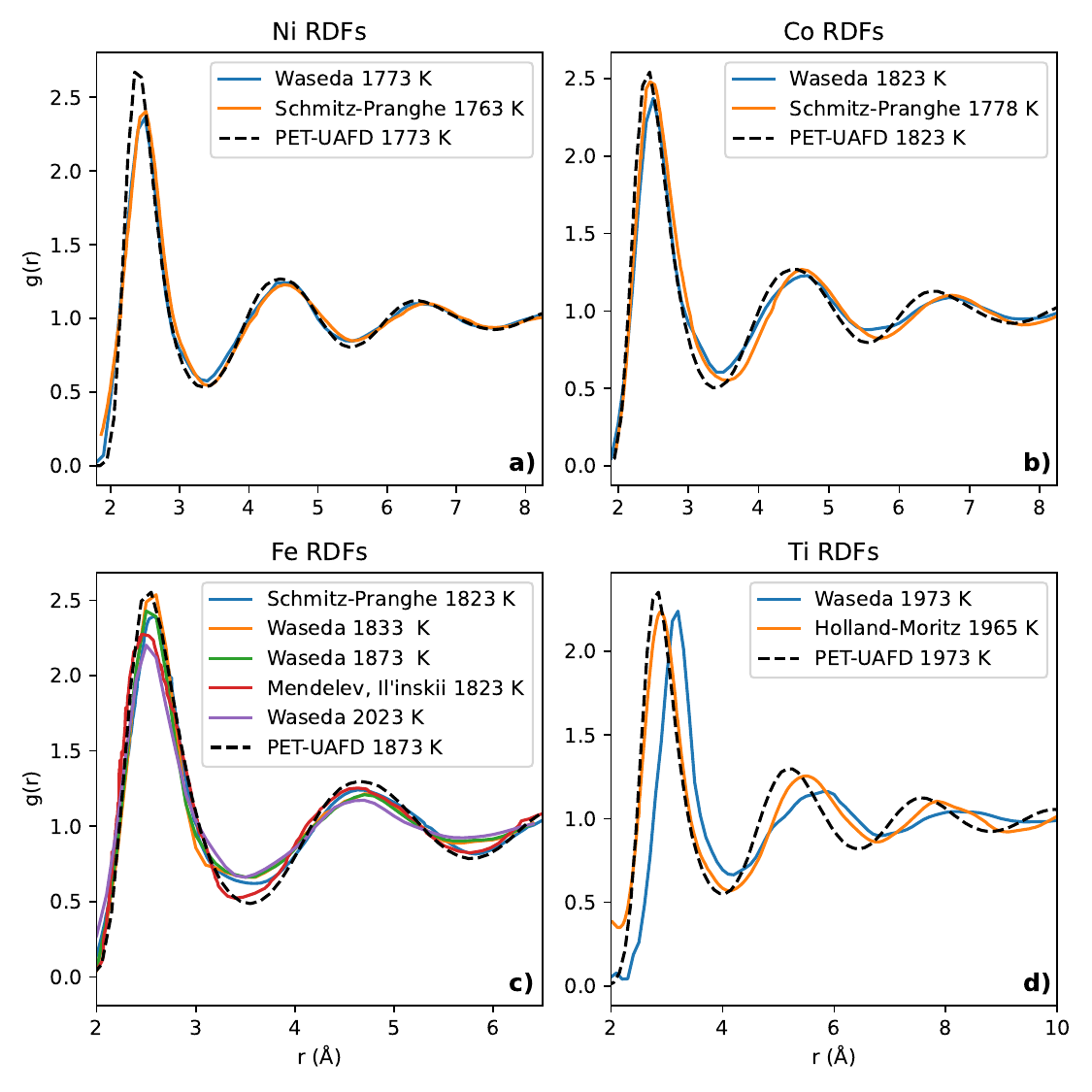}
    \caption{Comparing various experimental sources for selected liquid metals: a) nickel, b) cobalt, c) iron and d) titanium. For comparison, we plot the \petuafd predictions as a dashed black line. }
    \label{fig:si_experimental_uncertainties}
\end{figure}

The experimental values presented in this work are affected by a degree of uncertainty that is difficult to assess given the historical nature of the experimental data. In general, the database we assembled spans data points measured between the 1970s and 2010s. Measurements were performed with both neutron and x-ray scattering instruments. Ref.~\citenum{schmitz-prangheNotizenRontgenbeugungsuntersuchungenEisen1970}, for example, reports measurement uncertainties in the experimental temperatures of around $\pm 5$ K. We expect that these uncertainties are representative for other measurements and should not affect the measured radial distribution functions significantly. Waseda himself notes that \textit{the quantitative accuracy is inferior to that of other measurements [sic]} for his measurements of liquid V, Cr, Mn, Na (only at 823~K), and Fe (only at 2023~K). To the best of our knowledge, no other experimental sources exist for liquid Sc, V, Cr, and Mn, and we decided to include these data points in the database in order to display the simulated RDFs of all first-row liquid transition metals, although the reference data are of
unknown quality. In Figure~\ref{fig:si_experimental_uncertainties} we compare different experimental references of radial distribution functions from selected liquid elements, namely nickel, cobalt, iron and titanium. Here nickel and cobalt are representative of many samples in the database, for which we found several experimental sources in good agreement. We compare measurements from Waseda~\cite{waseda1980structure} and Schmitz-Pranghe~\cite{schmitz-prangheNotizenRontgenbeugungsuntersuchungenEisen1970}. The deviations remain small compared to the prediction errors made by \petuafd against experiment. Iron and titanium are cases for which conflicting experimental sources were found. 
For iron we compare experimental sources from Schmitz-Pranghe \cite{schmitz-prangheNotizenRontgenbeugungsuntersuchungenEisen1970} with those from Waseda \cite{waseda1980structure} at 1833 K, 1873 K, and 2023 K, and measurements from Il'inskii \cite{ilinskiiStructureLiquidFe2002}. The comparison between the various measurements indicates a significant deviation between the measurements of Ref.~\cite{ilinskiiStructureLiquidFe2002} and Refs.~\cite{schmitz-prangheNotizenRontgenbeugungsuntersuchungenEisen1970, waseda1980structure}.  Comparing this deviation with the temperature dependence of the RDFs from Waseda, strongly suggests that these deviations cannot be attributed solely to uncertainty in measurement conditions. The exact origins of these deviations remain elusive. In the case of Il'inskii, the authors mention potential contamination with iron oxide, but also stress that this should only affect the position of the first peak, and not other features of the RDF, which stands in contrast to the deviations we observe between the various sources which instead affect the depth of the well between the first two coordination shells. We choose to visualize the data of Il'inskii alongside the \petuafd predictions in the main text, as these data were measured using an instrument with
better low-Q resolution. Titanium is another case in which we found strongly conflicting sources, namely references~\cite{waseda1980structure} and \cite{holland-moritzShortrangeOrderStable2007}. Reports in the literature call Waseda's data \cite{waseda1980structure} \textit{somewhat unreliable [sic]} \cite{delrioFirstPrinciplesDetermination2020}. Ref.~\cite{delrioFirstPrinciplesDetermination2020} notes phase shift errors between their AIMD simulations and Waseda's data for titanium, scandium and vanadium. 
Considering that we observe similar phase shifts for liquid vanadium and scandium, this suggests that these samples are probably affected by a similar issue. We note that the \uafd prediction errors of the liquid densities of Sc and Ti are not particularly elevated - a large phase shift against experimental references in the computed RDF would almost certainly cause large density prediction errors. Given these observations, we choose to show the experimental reference data of Holland-Moritz alongside the \petuafd predictions in the main text and mark vanadium and scandium as potentially affected by measurement errors. However, it should be noted that the remaining discrepancy between \petuafd prediction of the liquid titanium RDF and the data from Holland-Moritz is most probably due to prediction errors of \petuafd, and therefore marks a case in which \petuafd uncertainty estimates are overconfident.

The measurements of liquid metal densities are also affected by measurement errors, although comparison of different sources reveals no major outliers, except for the density of liquid vanadium. After reanalyzing various experimental sources, Ntonti et
al. \cite{ntontiReferenceCorrelationsDensity2024} find density values for liquid vanadium that differ by 0.2 g/cm$^3$ from those published by Waseda. We choose to present the values from Ntonti et al. in this manuscript, as they appear to be in better agreement with newer primary sources that report density values around 5.5 g/cm$^3$.


\begin{thebibliography}{63}%
\makeatletter
\providecommand \@ifxundefined [1]{%
 \@ifx{#1\undefined}
}%
\providecommand \@ifnum [1]{%
 \ifnum #1\expandafter \@firstoftwo
 \else \expandafter \@secondoftwo
 \fi
}%
\providecommand \@ifx [1]{%
 \ifx #1\expandafter \@firstoftwo
 \else \expandafter \@secondoftwo
 \fi
}%
\providecommand \natexlab [1]{#1}%
\providecommand \enquote  [1]{``#1''}%
\providecommand \bibnamefont  [1]{#1}%
\providecommand \bibfnamefont [1]{#1}%
\providecommand \citenamefont [1]{#1}%
\providecommand \href@noop [0]{\@secondoftwo}%
\providecommand \href [0]{\begingroup \@sanitize@url \@href}%
\providecommand \@href[1]{\@@startlink{#1}\@@href}%
\providecommand \@@href[1]{\endgroup#1\@@endlink}%
\providecommand \@sanitize@url [0]{\catcode `\\12\catcode `\$12\catcode
  `\&12\catcode `\#12\catcode `\^12\catcode `\_12\catcode `\%12\relax}%
\providecommand \@@startlink[1]{}%
\providecommand \@@endlink[0]{}%
\providecommand \url  [0]{\begingroup\@sanitize@url \@url }%
\providecommand \@url [1]{\endgroup\@href {#1}{\urlprefix }}%
\providecommand \urlprefix  [0]{URL }%
\providecommand \Eprint [0]{\href }%
\providecommand \doibase [0]{https://doi.org/}%
\providecommand \selectlanguage [0]{\@gobble}%
\providecommand \bibinfo  [0]{\@secondoftwo}%
\providecommand \bibfield  [0]{\@secondoftwo}%
\providecommand \translation [1]{[#1]}%
\providecommand \BibitemOpen [0]{}%
\providecommand \bibitemStop [0]{}%
\providecommand \bibitemNoStop [0]{.\EOS\space}%
\providecommand \EOS [0]{\spacefactor3000\relax}%
\providecommand \BibitemShut  [1]{\csname bibitem#1\endcsname}%
\let\auto@bib@innerbib\@empty
\bibitem [{\citenamefont {Hafner}\ \emph {et~al.}(2006)\citenamefont {Hafner},
  \citenamefont {Wolverton},\ and\ \citenamefont
  {Ceder}}]{hafnerComputationalMaterialsDesign2006}%
  \BibitemOpen
  \bibfield  {author} {\bibinfo {author} {\bibfnamefont {J.}~\bibnamefont
  {Hafner}}, \bibinfo {author} {\bibfnamefont {C.}~\bibnamefont {Wolverton}},\
  and\ \bibinfo {author} {\bibfnamefont {G.}~\bibnamefont {Ceder}},\ }\bibfield
   {title} {\bibinfo {title} {Toward {{Computational Materials Design}}: {{The
  Impact}} of {{Density Functional Theory}} on {{Materials Research}}},\ }\href
  {https://doi.org/10.1557/mrs2006.174} {\bibfield  {journal} {\bibinfo
  {journal} {MRS Bulletin}\ }\textbf {\bibinfo {volume} {31}},\ \bibinfo
  {pages} {659} (\bibinfo {year} {2006})}\BibitemShut {NoStop}%
\bibitem [{\citenamefont {Burke}(2012)}]{burk12jcp}%
  \BibitemOpen
  \bibfield  {author} {\bibinfo {author} {\bibfnamefont {K.}~\bibnamefont
  {Burke}},\ }\bibfield  {title} {\bibinfo {title} {Perspective on density
  functional theory.},\ }\href {https://doi.org/10.1063/1.4704546} {\bibfield
  {journal} {\bibinfo  {journal} {The Journal of chemical physics}\ }\textbf
  {\bibinfo {volume} {136}},\ \bibinfo {pages} {150901} (\bibinfo {year}
  {2012})},\ \Eprint {https://arxiv.org/abs/22519306} {22519306} \BibitemShut
  {NoStop}%
\bibitem [{\citenamefont {Chen}\ and\ \citenamefont
  {Ong}(2022)}]{chenUniversalGraphDeep2022}%
  \BibitemOpen
  \bibfield  {author} {\bibinfo {author} {\bibfnamefont {C.}~\bibnamefont
  {Chen}}\ and\ \bibinfo {author} {\bibfnamefont {S.~P.}\ \bibnamefont {Ong}},\
  }\bibfield  {title} {\bibinfo {title} {A universal graph deep learning
  interatomic potential for the periodic table},\ }\href
  {https://doi.org/10.1038/s43588-022-00349-3} {\bibfield  {journal} {\bibinfo
  {journal} {Nature Computational Science}\ }\textbf {\bibinfo {volume} {2}},\
  \bibinfo {pages} {718} (\bibinfo {year} {2022})}\BibitemShut {NoStop}%
\bibitem [{\citenamefont {Deng}\ \emph {et~al.}(2023)\citenamefont {Deng},
  \citenamefont {Zhong}, \citenamefont {Jun}, \citenamefont {Riebesell},
  \citenamefont {Han}, \citenamefont {Bartel},\ and\ \citenamefont
  {Ceder}}]{dengCHGNetPretrainedUniversal2023}%
  \BibitemOpen
  \bibfield  {author} {\bibinfo {author} {\bibfnamefont {B.}~\bibnamefont
  {Deng}}, \bibinfo {author} {\bibfnamefont {P.}~\bibnamefont {Zhong}},
  \bibinfo {author} {\bibfnamefont {K.}~\bibnamefont {Jun}}, \bibinfo {author}
  {\bibfnamefont {J.}~\bibnamefont {Riebesell}}, \bibinfo {author}
  {\bibfnamefont {K.}~\bibnamefont {Han}}, \bibinfo {author} {\bibfnamefont
  {C.~J.}\ \bibnamefont {Bartel}},\ and\ \bibinfo {author} {\bibfnamefont
  {G.}~\bibnamefont {Ceder}},\ }\bibfield  {title} {\bibinfo {title}
  {{{CHGNet}} as a pretrained universal neural network potential for
  charge-informed atomistic modelling},\ }\href
  {https://doi.org/10.1038/s42256-023-00716-3} {\bibfield  {journal} {\bibinfo
  {journal} {Nature Machine Intelligence}\ }\textbf {\bibinfo {volume} {5}},\
  \bibinfo {pages} {1031} (\bibinfo {year} {2023})}\BibitemShut {NoStop}%
\bibitem [{\citenamefont {Batatia}\ \emph {et~al.}(2025)\citenamefont
  {Batatia}, \citenamefont {Benner}, \citenamefont {Chiang}, \citenamefont
  {Elena}, \citenamefont {Kov{\'a}cs}, \citenamefont {Riebesell}, \citenamefont
  {Advincula}, \citenamefont {Asta}, \citenamefont {Avaylon}, \citenamefont
  {Baldwin}, \citenamefont {Berger}, \citenamefont {Bernstein}, \citenamefont
  {Bhowmik}, \citenamefont {Bigi}, \citenamefont {Blau}, \citenamefont {C{\u
  a}rare}, \citenamefont {Ceriotti}, \citenamefont {Chong}, \citenamefont
  {Darby}, \citenamefont {De}, \citenamefont {Della~Pia}, \citenamefont
  {Deringer}, \citenamefont {Elijo{\v s}ius}, \citenamefont {{El-Machachi}},
  \citenamefont {Fako}, \citenamefont {Falcioni}, \citenamefont {Ferrari},
  \citenamefont {Gardner}, \citenamefont {Gawkowski}, \citenamefont
  {{Genreith-Schriever}}, \citenamefont {George}, \citenamefont {Goodall},
  \citenamefont {Grandel}, \citenamefont {Grey}, \citenamefont {Grigorev},
  \citenamefont {Han}, \citenamefont {Handley}, \citenamefont {Heenen},
  \citenamefont {Hermansson}, \citenamefont {Ho}, \citenamefont {Hofmann},
  \citenamefont {Holm}, \citenamefont {Jaafar}, \citenamefont {Jakob},
  \citenamefont {Jung}, \citenamefont {Kapil}, \citenamefont {Kaplan},
  \citenamefont {Karimitari}, \citenamefont {Kermode}, \citenamefont {Kourtis},
  \citenamefont {Kroupa}, \citenamefont {Kullgren}, \citenamefont {Kuner},
  \citenamefont {Kuryla}, \citenamefont {Liepuoniute}, \citenamefont {Lin},
  \citenamefont {Margraf}, \citenamefont {Magd{\u a}u}, \citenamefont
  {Michaelides}, \citenamefont {Moore}, \citenamefont {Naik}, \citenamefont
  {Niblett}, \citenamefont {Norwood}, \citenamefont {O'Neill}, \citenamefont
  {Ortner}, \citenamefont {Persson}, \citenamefont {Reuter}, \citenamefont
  {Rosen}, \citenamefont {Rosset}, \citenamefont {Schaaf}, \citenamefont
  {Schran}, \citenamefont {Shi}, \citenamefont {Sivonxay}, \citenamefont
  {Stenczel}, \citenamefont {Sutton}, \citenamefont {Svahn}, \citenamefont
  {Swinburne}, \citenamefont {Tilly}, \citenamefont {Van Der~Oord},
  \citenamefont {Vargas}, \citenamefont {{Varga-Umbrich}}, \citenamefont
  {Vegge}, \citenamefont {Vondr{\'a}k}, \citenamefont {Wang}, \citenamefont
  {Witt}, \citenamefont {Wolf}, \citenamefont {Zills},\ and\ \citenamefont
  {Cs{\'a}nyi}}]{bata+25jcp}%
  \BibitemOpen
  \bibfield  {author} {\bibinfo {author} {\bibfnamefont {I.}~\bibnamefont
  {Batatia}}, \bibinfo {author} {\bibfnamefont {P.}~\bibnamefont {Benner}},
  \bibinfo {author} {\bibfnamefont {Y.}~\bibnamefont {Chiang}}, \bibinfo
  {author} {\bibfnamefont {A.~M.}\ \bibnamefont {Elena}}, \bibinfo {author}
  {\bibfnamefont {D.~P.}\ \bibnamefont {Kov{\'a}cs}}, \bibinfo {author}
  {\bibfnamefont {J.}~\bibnamefont {Riebesell}}, \bibinfo {author}
  {\bibfnamefont {X.~R.}\ \bibnamefont {Advincula}}, \bibinfo {author}
  {\bibfnamefont {M.}~\bibnamefont {Asta}}, \bibinfo {author} {\bibfnamefont
  {M.}~\bibnamefont {Avaylon}}, \bibinfo {author} {\bibfnamefont {W.~J.}\
  \bibnamefont {Baldwin}}, \bibinfo {author} {\bibfnamefont {F.}~\bibnamefont
  {Berger}}, \bibinfo {author} {\bibfnamefont {N.}~\bibnamefont {Bernstein}},
  \bibinfo {author} {\bibfnamefont {A.}~\bibnamefont {Bhowmik}}, \bibinfo
  {author} {\bibfnamefont {F.}~\bibnamefont {Bigi}}, \bibinfo {author}
  {\bibfnamefont {S.~M.}\ \bibnamefont {Blau}}, \bibinfo {author}
  {\bibfnamefont {V.}~\bibnamefont {C{\u a}rare}}, \bibinfo {author}
  {\bibfnamefont {M.}~\bibnamefont {Ceriotti}}, \bibinfo {author}
  {\bibfnamefont {S.}~\bibnamefont {Chong}}, \bibinfo {author} {\bibfnamefont
  {J.~P.}\ \bibnamefont {Darby}}, \bibinfo {author} {\bibfnamefont
  {S.}~\bibnamefont {De}}, \bibinfo {author} {\bibfnamefont {F.}~\bibnamefont
  {Della~Pia}}, \bibinfo {author} {\bibfnamefont {V.~L.}\ \bibnamefont
  {Deringer}}, \bibinfo {author} {\bibfnamefont {R.}~\bibnamefont {Elijo{\v
  s}ius}}, \bibinfo {author} {\bibfnamefont {Z.}~\bibnamefont {{El-Machachi}}},
  \bibinfo {author} {\bibfnamefont {E.}~\bibnamefont {Fako}}, \bibinfo {author}
  {\bibfnamefont {F.}~\bibnamefont {Falcioni}}, \bibinfo {author}
  {\bibfnamefont {A.~C.}\ \bibnamefont {Ferrari}}, \bibinfo {author}
  {\bibfnamefont {J.~L.~A.}\ \bibnamefont {Gardner}}, \bibinfo {author}
  {\bibfnamefont {M.~J.}\ \bibnamefont {Gawkowski}}, \bibinfo {author}
  {\bibfnamefont {A.}~\bibnamefont {{Genreith-Schriever}}}, \bibinfo {author}
  {\bibfnamefont {J.}~\bibnamefont {George}}, \bibinfo {author} {\bibfnamefont
  {R.~E.~A.}\ \bibnamefont {Goodall}}, \bibinfo {author} {\bibfnamefont
  {J.}~\bibnamefont {Grandel}}, \bibinfo {author} {\bibfnamefont {C.~P.}\
  \bibnamefont {Grey}}, \bibinfo {author} {\bibfnamefont {P.}~\bibnamefont
  {Grigorev}}, \bibinfo {author} {\bibfnamefont {S.}~\bibnamefont {Han}},
  \bibinfo {author} {\bibfnamefont {W.}~\bibnamefont {Handley}}, \bibinfo
  {author} {\bibfnamefont {H.~H.}\ \bibnamefont {Heenen}}, \bibinfo {author}
  {\bibfnamefont {K.}~\bibnamefont {Hermansson}}, \bibinfo {author}
  {\bibfnamefont {C.~H.}\ \bibnamefont {Ho}}, \bibinfo {author} {\bibfnamefont
  {S.}~\bibnamefont {Hofmann}}, \bibinfo {author} {\bibfnamefont
  {C.}~\bibnamefont {Holm}}, \bibinfo {author} {\bibfnamefont {J.}~\bibnamefont
  {Jaafar}}, \bibinfo {author} {\bibfnamefont {K.~S.}\ \bibnamefont {Jakob}},
  \bibinfo {author} {\bibfnamefont {H.}~\bibnamefont {Jung}}, \bibinfo {author}
  {\bibfnamefont {V.}~\bibnamefont {Kapil}}, \bibinfo {author} {\bibfnamefont
  {A.~D.}\ \bibnamefont {Kaplan}}, \bibinfo {author} {\bibfnamefont
  {N.}~\bibnamefont {Karimitari}}, \bibinfo {author} {\bibfnamefont {J.~R.}\
  \bibnamefont {Kermode}}, \bibinfo {author} {\bibfnamefont {P.}~\bibnamefont
  {Kourtis}}, \bibinfo {author} {\bibfnamefont {N.}~\bibnamefont {Kroupa}},
  \bibinfo {author} {\bibfnamefont {J.}~\bibnamefont {Kullgren}}, \bibinfo
  {author} {\bibfnamefont {M.~C.}\ \bibnamefont {Kuner}}, \bibinfo {author}
  {\bibfnamefont {D.}~\bibnamefont {Kuryla}}, \bibinfo {author} {\bibfnamefont
  {G.}~\bibnamefont {Liepuoniute}}, \bibinfo {author} {\bibfnamefont
  {C.}~\bibnamefont {Lin}}, \bibinfo {author} {\bibfnamefont {J.~T.}\
  \bibnamefont {Margraf}}, \bibinfo {author} {\bibfnamefont {I.-B.}\
  \bibnamefont {Magd{\u a}u}}, \bibinfo {author} {\bibfnamefont
  {A.}~\bibnamefont {Michaelides}}, \bibinfo {author} {\bibfnamefont {J.~H.}\
  \bibnamefont {Moore}}, \bibinfo {author} {\bibfnamefont {A.~A.}\ \bibnamefont
  {Naik}}, \bibinfo {author} {\bibfnamefont {S.~P.}\ \bibnamefont {Niblett}},
  \bibinfo {author} {\bibfnamefont {S.~W.}\ \bibnamefont {Norwood}}, \bibinfo
  {author} {\bibfnamefont {N.}~\bibnamefont {O'Neill}}, \bibinfo {author}
  {\bibfnamefont {C.}~\bibnamefont {Ortner}}, \bibinfo {author} {\bibfnamefont
  {K.~A.}\ \bibnamefont {Persson}}, \bibinfo {author} {\bibfnamefont
  {K.}~\bibnamefont {Reuter}}, \bibinfo {author} {\bibfnamefont {A.~S.}\
  \bibnamefont {Rosen}}, \bibinfo {author} {\bibfnamefont {L.~A.~M.}\
  \bibnamefont {Rosset}}, \bibinfo {author} {\bibfnamefont {L.~L.}\
  \bibnamefont {Schaaf}}, \bibinfo {author} {\bibfnamefont {C.}~\bibnamefont
  {Schran}}, \bibinfo {author} {\bibfnamefont {B.~X.}\ \bibnamefont {Shi}},
  \bibinfo {author} {\bibfnamefont {E.}~\bibnamefont {Sivonxay}}, \bibinfo
  {author} {\bibfnamefont {T.~K.}\ \bibnamefont {Stenczel}}, \bibinfo {author}
  {\bibfnamefont {C.}~\bibnamefont {Sutton}}, \bibinfo {author} {\bibfnamefont
  {V.}~\bibnamefont {Svahn}}, \bibinfo {author} {\bibfnamefont {T.~D.}\
  \bibnamefont {Swinburne}}, \bibinfo {author} {\bibfnamefont {J.}~\bibnamefont
  {Tilly}}, \bibinfo {author} {\bibfnamefont {C.}~\bibnamefont {Van Der~Oord}},
  \bibinfo {author} {\bibfnamefont {S.}~\bibnamefont {Vargas}}, \bibinfo
  {author} {\bibfnamefont {E.}~\bibnamefont {{Varga-Umbrich}}}, \bibinfo
  {author} {\bibfnamefont {T.}~\bibnamefont {Vegge}}, \bibinfo {author}
  {\bibfnamefont {M.}~\bibnamefont {Vondr{\'a}k}}, \bibinfo {author}
  {\bibfnamefont {Y.}~\bibnamefont {Wang}}, \bibinfo {author} {\bibfnamefont
  {W.~C.}\ \bibnamefont {Witt}}, \bibinfo {author} {\bibfnamefont
  {T.}~\bibnamefont {Wolf}}, \bibinfo {author} {\bibfnamefont {F.}~\bibnamefont
  {Zills}},\ and\ \bibinfo {author} {\bibfnamefont {G.}~\bibnamefont
  {Cs{\'a}nyi}},\ }\bibfield  {title} {\bibinfo {title} {A foundation model for
  atomistic materials chemistry},\ }\href {https://doi.org/10.1063/5.0297006}
  {\bibfield  {journal} {\bibinfo  {journal} {The Journal of Chemical Physics}\
  }\textbf {\bibinfo {volume} {163}},\ \bibinfo {pages} {184110} (\bibinfo
  {year} {2025})}\BibitemShut {NoStop}%
\bibitem [{\citenamefont {Wood}\ \emph {et~al.}(2026)\citenamefont {Wood},
  \citenamefont {Dzamba}, \citenamefont {Fu}, \citenamefont {Gao},
  \citenamefont {Shuaibi}, \citenamefont {{Barroso-Luque}}, \citenamefont
  {Abdelmaqsoud}, \citenamefont {Gharakhanyan}, \citenamefont {Kitchin},
  \citenamefont {Levine}, \citenamefont {Michel}, \citenamefont {Sriram},
  \citenamefont {Cohen}, \citenamefont {Das}, \citenamefont {Rizvi},
  \citenamefont {Sahoo}, \citenamefont {Ulissi},\ and\ \citenamefont
  {Zitnick}}]{woodUMAFamilyUniversal2026}%
  \BibitemOpen
  \bibfield  {author} {\bibinfo {author} {\bibfnamefont {B.~M.}\ \bibnamefont
  {Wood}}, \bibinfo {author} {\bibfnamefont {M.}~\bibnamefont {Dzamba}},
  \bibinfo {author} {\bibfnamefont {X.}~\bibnamefont {Fu}}, \bibinfo {author}
  {\bibfnamefont {M.}~\bibnamefont {Gao}}, \bibinfo {author} {\bibfnamefont
  {M.}~\bibnamefont {Shuaibi}}, \bibinfo {author} {\bibfnamefont
  {L.}~\bibnamefont {{Barroso-Luque}}}, \bibinfo {author} {\bibfnamefont
  {K.}~\bibnamefont {Abdelmaqsoud}}, \bibinfo {author} {\bibfnamefont
  {V.}~\bibnamefont {Gharakhanyan}}, \bibinfo {author} {\bibfnamefont {J.~R.}\
  \bibnamefont {Kitchin}}, \bibinfo {author} {\bibfnamefont {D.~S.}\
  \bibnamefont {Levine}}, \bibinfo {author} {\bibfnamefont {K.}~\bibnamefont
  {Michel}}, \bibinfo {author} {\bibfnamefont {A.}~\bibnamefont {Sriram}},
  \bibinfo {author} {\bibfnamefont {T.}~\bibnamefont {Cohen}}, \bibinfo
  {author} {\bibfnamefont {A.}~\bibnamefont {Das}}, \bibinfo {author}
  {\bibfnamefont {A.}~\bibnamefont {Rizvi}}, \bibinfo {author} {\bibfnamefont
  {S.~J.}\ \bibnamefont {Sahoo}}, \bibinfo {author} {\bibfnamefont {Z.~W.}\
  \bibnamefont {Ulissi}},\ and\ \bibinfo {author} {\bibfnamefont {C.~L.}\
  \bibnamefont {Zitnick}},\ }\href {https://doi.org/10.48550/arXiv.2506.23971}
  {\bibinfo {title} {{{UMA}}: {{A Family}} of {{Universal Models}} for
  {{Atoms}}}} (\bibinfo {year} {2026}),\ \Eprint
  {https://arxiv.org/abs/2506.23971} {arXiv:2506.23971 [cs]} \BibitemShut
  {NoStop}%
\bibitem [{\citenamefont {Mazitov}\ \emph
  {et~al.}(2025{\natexlab{a}})\citenamefont {Mazitov}, \citenamefont {Bigi},
  \citenamefont {Kellner}, \citenamefont {Pegolo}, \citenamefont {Tisi},
  \citenamefont {Fraux}, \citenamefont {Pozdnyakov}, \citenamefont {Loche},\
  and\ \citenamefont {Ceriotti}}]{mazi+25ncomm}%
  \BibitemOpen
  \bibfield  {author} {\bibinfo {author} {\bibfnamefont {A.}~\bibnamefont
  {Mazitov}}, \bibinfo {author} {\bibfnamefont {F.}~\bibnamefont {Bigi}},
  \bibinfo {author} {\bibfnamefont {M.}~\bibnamefont {Kellner}}, \bibinfo
  {author} {\bibfnamefont {P.}~\bibnamefont {Pegolo}}, \bibinfo {author}
  {\bibfnamefont {D.}~\bibnamefont {Tisi}}, \bibinfo {author} {\bibfnamefont
  {G.}~\bibnamefont {Fraux}}, \bibinfo {author} {\bibfnamefont
  {S.}~\bibnamefont {Pozdnyakov}}, \bibinfo {author} {\bibfnamefont
  {P.}~\bibnamefont {Loche}},\ and\ \bibinfo {author} {\bibfnamefont
  {M.}~\bibnamefont {Ceriotti}},\ }\bibfield  {title} {\bibinfo {title}
  {{{PET-MAD}} as a lightweight universal interatomic potential for advanced
  materials modeling},\ }\href {https://doi.org/10.1038/s41467-025-65662-7}
  {\bibfield  {journal} {\bibinfo  {journal} {Nat Commun}\ }\textbf {\bibinfo
  {volume} {16}},\ \bibinfo {pages} {10653} (\bibinfo {year}
  {2025}{\natexlab{a}})}\BibitemShut {NoStop}%
\bibitem [{\citenamefont {Musil}\ \emph {et~al.}(2019)\citenamefont {Musil},
  \citenamefont {Willatt}, \citenamefont {Langovoy},\ and\ \citenamefont
  {Ceriotti}}]{musi+19jctc}%
  \BibitemOpen
  \bibfield  {author} {\bibinfo {author} {\bibfnamefont {F.}~\bibnamefont
  {Musil}}, \bibinfo {author} {\bibfnamefont {M.~J.}\ \bibnamefont {Willatt}},
  \bibinfo {author} {\bibfnamefont {M.~A.}\ \bibnamefont {Langovoy}},\ and\
  \bibinfo {author} {\bibfnamefont {M.}~\bibnamefont {Ceriotti}},\ }\bibfield
  {title} {\bibinfo {title} {Fast and {{Accurate Uncertainty Estimation}} in
  {{Chemical Machine Learning}}},\ }\href
  {https://doi.org/10.1021/acs.jctc.8b00959} {\bibfield  {journal} {\bibinfo
  {journal} {Journal of Chemical Theory and Computation}\ }\textbf {\bibinfo
  {volume} {15}},\ \bibinfo {pages} {906} (\bibinfo {year} {2019})}\BibitemShut
  {NoStop}%
\bibitem [{\citenamefont {Imbalzano}\ \emph {et~al.}(2021)\citenamefont
  {Imbalzano}, \citenamefont {Zhuang}, \citenamefont {Kapil}, \citenamefont
  {Rossi}, \citenamefont {Engel}, \citenamefont {Grasselli},\ and\
  \citenamefont {Ceriotti}}]{imba+21jcp}%
  \BibitemOpen
  \bibfield  {author} {\bibinfo {author} {\bibfnamefont {G.}~\bibnamefont
  {Imbalzano}}, \bibinfo {author} {\bibfnamefont {Y.}~\bibnamefont {Zhuang}},
  \bibinfo {author} {\bibfnamefont {V.}~\bibnamefont {Kapil}}, \bibinfo
  {author} {\bibfnamefont {K.}~\bibnamefont {Rossi}}, \bibinfo {author}
  {\bibfnamefont {E.~A.}\ \bibnamefont {Engel}}, \bibinfo {author}
  {\bibfnamefont {F.}~\bibnamefont {Grasselli}},\ and\ \bibinfo {author}
  {\bibfnamefont {M.}~\bibnamefont {Ceriotti}},\ }\bibfield  {title} {\bibinfo
  {title} {Uncertainty estimation for molecular dynamics and sampling},\ }\href
  {https://doi.org/10.1063/5.0036522} {\bibfield  {journal} {\bibinfo
  {journal} {J. Chem. Phys.}\ }\textbf {\bibinfo {volume} {154}},\ \bibinfo
  {pages} {074102} (\bibinfo {year} {2021})}\BibitemShut {NoStop}%
\bibitem [{\citenamefont {Pernot}(2022)}]{pern22jcp}%
  \BibitemOpen
  \bibfield  {author} {\bibinfo {author} {\bibfnamefont {P.}~\bibnamefont
  {Pernot}},\ }\bibfield  {title} {\bibinfo {title} {The long road to
  calibrated prediction uncertainty in computational chemistry},\ }\href
  {https://doi.org/10.1063/5.0084302} {\bibfield  {journal} {\bibinfo
  {journal} {The Journal of Chemical Physics}\ }\textbf {\bibinfo {volume}
  {156}},\ \bibinfo {pages} {114109} (\bibinfo {year} {2022})}\BibitemShut
  {NoStop}%
\bibitem [{\citenamefont {Kellner}\ and\ \citenamefont
  {Ceriotti}(2024)}]{kell-ceri24mlst}%
  \BibitemOpen
  \bibfield  {author} {\bibinfo {author} {\bibfnamefont {M.}~\bibnamefont
  {Kellner}}\ and\ \bibinfo {author} {\bibfnamefont {M.}~\bibnamefont
  {Ceriotti}},\ }\bibfield  {title} {\bibinfo {title} {Uncertainty
  quantification by direct propagation of shallow ensembles},\ }\href
  {https://doi.org/10.1088/2632-2153/ad594a} {\bibfield  {journal} {\bibinfo
  {journal} {Mach. Learn.: Sci. Technol.}\ }\textbf {\bibinfo {volume} {5}},\
  \bibinfo {pages} {035006} (\bibinfo {year} {2024})}\BibitemShut {NoStop}%
\bibitem [{\citenamefont {Bigi}\ \emph {et~al.}(2024)\citenamefont {Bigi},
  \citenamefont {Chong}, \citenamefont {Ceriotti},\ and\ \citenamefont
  {Grasselli}}]{bigi+24mlst}%
  \BibitemOpen
  \bibfield  {author} {\bibinfo {author} {\bibfnamefont {F.}~\bibnamefont
  {Bigi}}, \bibinfo {author} {\bibfnamefont {S.}~\bibnamefont {Chong}},
  \bibinfo {author} {\bibfnamefont {M.}~\bibnamefont {Ceriotti}},\ and\
  \bibinfo {author} {\bibfnamefont {F.}~\bibnamefont {Grasselli}},\ }\bibfield
  {title} {\bibinfo {title} {A prediction rigidity formalism for low-cost
  uncertainties in trained neural networks},\ }\href
  {https://doi.org/10.1088/2632-2153/ad805f} {\bibfield  {journal} {\bibinfo
  {journal} {Mach. Learn.: Sci. Technol.}\ }\textbf {\bibinfo {volume} {5}},\
  \bibinfo {pages} {045018} (\bibinfo {year} {2024})}\BibitemShut {NoStop}%
\bibitem [{\citenamefont {Zhu}\ \emph {et~al.}(2023)\citenamefont {Zhu},
  \citenamefont {Batzner}, \citenamefont {Musaelian},\ and\ \citenamefont
  {Kozinsky}}]{zhuFastUncertaintyEstimates2023}%
  \BibitemOpen
  \bibfield  {author} {\bibinfo {author} {\bibfnamefont {A.}~\bibnamefont
  {Zhu}}, \bibinfo {author} {\bibfnamefont {S.}~\bibnamefont {Batzner}},
  \bibinfo {author} {\bibfnamefont {A.}~\bibnamefont {Musaelian}},\ and\
  \bibinfo {author} {\bibfnamefont {B.}~\bibnamefont {Kozinsky}},\ }\bibfield
  {title} {\bibinfo {title} {Fast uncertainty estimates in deep learning
  interatomic potentials},\ }\href {https://doi.org/10.1063/5.0136574}
  {\bibfield  {journal} {\bibinfo  {journal} {The Journal of Chemical Physics}\
  }\textbf {\bibinfo {volume} {158}},\ \bibinfo {pages} {164111} (\bibinfo
  {year} {2023})}\BibitemShut {NoStop}%
\bibitem [{\citenamefont {Perez}\ \emph {et~al.}(2025)\citenamefont {Perez},
  \citenamefont {Subramanyam}, \citenamefont {Maliyov},\ and\ \citenamefont
  {Swinburne}}]{perezUncertaintyQuantificationMisspecified2025}%
  \BibitemOpen
  \bibfield  {author} {\bibinfo {author} {\bibfnamefont {D.}~\bibnamefont
  {Perez}}, \bibinfo {author} {\bibfnamefont {A.~P.~A.}\ \bibnamefont
  {Subramanyam}}, \bibinfo {author} {\bibfnamefont {I.}~\bibnamefont
  {Maliyov}},\ and\ \bibinfo {author} {\bibfnamefont {T.~D.}\ \bibnamefont
  {Swinburne}},\ }\bibfield  {title} {\bibinfo {title} {Uncertainty
  quantification for misspecified machine learned interatomic potentials},\
  }\href {https://doi.org/10.1038/s41524-025-01758-4} {\bibfield  {journal}
  {\bibinfo  {journal} {npj Computational Materials}\ }\textbf {\bibinfo
  {volume} {11}},\ \bibinfo {pages} {263} (\bibinfo {year} {2025})}\BibitemShut
  {NoStop}%
\bibitem [{\citenamefont {Mortensen}\ \emph {et~al.}(2005)\citenamefont
  {Mortensen}, \citenamefont {Kaasbjerg}, \citenamefont {Frederiksen},
  \citenamefont {N{\o}rskov}, \citenamefont {Sethna},\ and\ \citenamefont
  {Jacobsen}}]{mortensenBayesianErrorEstimation2005}%
  \BibitemOpen
  \bibfield  {author} {\bibinfo {author} {\bibfnamefont {J.~J.}\ \bibnamefont
  {Mortensen}}, \bibinfo {author} {\bibfnamefont {K.}~\bibnamefont
  {Kaasbjerg}}, \bibinfo {author} {\bibfnamefont {S.~L.}\ \bibnamefont
  {Frederiksen}}, \bibinfo {author} {\bibfnamefont {J.~K.}\ \bibnamefont
  {N{\o}rskov}}, \bibinfo {author} {\bibfnamefont {J.~P.}\ \bibnamefont
  {Sethna}},\ and\ \bibinfo {author} {\bibfnamefont {K.~W.}\ \bibnamefont
  {Jacobsen}},\ }\bibfield  {title} {\bibinfo {title} {Bayesian {{Error
  Estimation}} in {{Density-Functional Theory}}},\ }\href
  {https://doi.org/10.1103/PhysRevLett.95.216401} {\bibfield  {journal}
  {\bibinfo  {journal} {Physical Review Letters}\ }\textbf {\bibinfo {volume}
  {95}},\ \bibinfo {pages} {216401} (\bibinfo {year} {2005})}\BibitemShut
  {NoStop}%
\bibitem [{\citenamefont {Wellendorff}\ \emph {et~al.}(2012)\citenamefont
  {Wellendorff}, \citenamefont {Lundgaard}, \citenamefont {M{\o}gelh{\o}j},
  \citenamefont {Petzold}, \citenamefont {Landis}, \citenamefont {N{\o}rskov},
  \citenamefont {Bligaard},\ and\ \citenamefont
  {Jacobsen}}]{wellendorffDensityFunctionalsSurface2012a}%
  \BibitemOpen
  \bibfield  {author} {\bibinfo {author} {\bibfnamefont {J.}~\bibnamefont
  {Wellendorff}}, \bibinfo {author} {\bibfnamefont {K.~T.}\ \bibnamefont
  {Lundgaard}}, \bibinfo {author} {\bibfnamefont {A.}~\bibnamefont
  {M{\o}gelh{\o}j}}, \bibinfo {author} {\bibfnamefont {V.}~\bibnamefont
  {Petzold}}, \bibinfo {author} {\bibfnamefont {D.~D.}\ \bibnamefont {Landis}},
  \bibinfo {author} {\bibfnamefont {J.~K.}\ \bibnamefont {N{\o}rskov}},
  \bibinfo {author} {\bibfnamefont {T.}~\bibnamefont {Bligaard}},\ and\
  \bibinfo {author} {\bibfnamefont {K.~W.}\ \bibnamefont {Jacobsen}},\
  }\bibfield  {title} {\bibinfo {title} {Density functionals for surface
  science: {{Exchange-correlation}} model development with {{Bayesian}} error
  estimation},\ }\href {https://doi.org/10.1103/PhysRevB.85.235149} {\bibfield
  {journal} {\bibinfo  {journal} {Physical Review B}\ }\textbf {\bibinfo
  {volume} {85}},\ \bibinfo {pages} {235149} (\bibinfo {year}
  {2012})}\BibitemShut {NoStop}%
\bibitem [{\citenamefont {Aldegunde}\ \emph {et~al.}(2016)\citenamefont
  {Aldegunde}, \citenamefont {Kermode},\ and\ \citenamefont
  {Zabaras}}]{aldegundeDevelopmentExchangeCorrelation2016}%
  \BibitemOpen
  \bibfield  {author} {\bibinfo {author} {\bibfnamefont {M.}~\bibnamefont
  {Aldegunde}}, \bibinfo {author} {\bibfnamefont {J.~R.}\ \bibnamefont
  {Kermode}},\ and\ \bibinfo {author} {\bibfnamefont {N.}~\bibnamefont
  {Zabaras}},\ }\bibfield  {title} {\bibinfo {title} {Development of an
  exchange--correlation functional with uncertainty quantification capabilities
  for density functional theory},\ }\href
  {https://doi.org/10.1016/j.jcp.2016.01.034} {\bibfield  {journal} {\bibinfo
  {journal} {Journal of Computational Physics}\ }\textbf {\bibinfo {volume}
  {311}},\ \bibinfo {pages} {173} (\bibinfo {year} {2016})}\BibitemShut
  {NoStop}%
\bibitem [{\citenamefont {Wellendorff}\ \emph {et~al.}(2014)\citenamefont
  {Wellendorff}, \citenamefont {Lundgaard}, \citenamefont {Jacobsen},\ and\
  \citenamefont {Bligaard}}]{wellendorffMBEEFAccurateSemilocal2014a}%
  \BibitemOpen
  \bibfield  {author} {\bibinfo {author} {\bibfnamefont {J.}~\bibnamefont
  {Wellendorff}}, \bibinfo {author} {\bibfnamefont {K.~T.}\ \bibnamefont
  {Lundgaard}}, \bibinfo {author} {\bibfnamefont {K.~W.}\ \bibnamefont
  {Jacobsen}},\ and\ \bibinfo {author} {\bibfnamefont {T.}~\bibnamefont
  {Bligaard}},\ }\bibfield  {title} {\bibinfo {title} {{{mBEEF}}: {{An}}
  accurate semi-local {{Bayesian}} error estimation density functional},\
  }\href {https://doi.org/10.1063/1.4870397} {\bibfield  {journal} {\bibinfo
  {journal} {The Journal of Chemical Physics}\ }\textbf {\bibinfo {volume}
  {140}},\ \bibinfo {pages} {144107} (\bibinfo {year} {2014})}\BibitemShut
  {NoStop}%
\bibitem [{\citenamefont {Hansen}\ \emph {et~al.}(2025)\citenamefont {Hansen},
  \citenamefont {Mortensen}, \citenamefont {Bligaard},\ and\ \citenamefont
  {Jacobsen}}]{hansenUncertaintyawareElectronicDensityfunctional2025}%
  \BibitemOpen
  \bibfield  {author} {\bibinfo {author} {\bibfnamefont {T.}~\bibnamefont
  {Hansen}}, \bibinfo {author} {\bibfnamefont {J.~J.}\ \bibnamefont
  {Mortensen}}, \bibinfo {author} {\bibfnamefont {T.}~\bibnamefont
  {Bligaard}},\ and\ \bibinfo {author} {\bibfnamefont {K.~W.}\ \bibnamefont
  {Jacobsen}},\ }\bibfield  {title} {\bibinfo {title} {Uncertainty-aware
  electronic density-functional distributions},\ }\href
  {https://doi.org/10.1103/yhly-wxhv} {\bibfield  {journal} {\bibinfo
  {journal} {Physical Review B}\ }\textbf {\bibinfo {volume} {112}},\ \bibinfo
  {pages} {075412} (\bibinfo {year} {2025})}\BibitemShut {NoStop}%
\bibitem [{\citenamefont {Pozdnyakov}\ and\ \citenamefont
  {Ceriotti}(2023)}]{pozd-ceri23nips}%
  \BibitemOpen
  \bibfield  {author} {\bibinfo {author} {\bibfnamefont {S.}~\bibnamefont
  {Pozdnyakov}}\ and\ \bibinfo {author} {\bibfnamefont {M.}~\bibnamefont
  {Ceriotti}},\ }\bibfield  {title} {\bibinfo {title} {Smooth, exact rotational
  symmetrization for deep learning on point clouds},\ }in\ \href@noop {} {\emph
  {\bibinfo {booktitle} {Advances in {{Neural Information Processing
  Systems}}}}},\ Vol.~\bibinfo {volume} {36}\ (\bibinfo  {publisher} {Curran
  Associates, Inc.},\ \bibinfo {year} {2023})\ pp.\ \bibinfo {pages}
  {79469--79501}\BibitemShut {NoStop}%
\bibitem [{\citenamefont {Bigi}\ \emph
  {et~al.}(2026{\natexlab{a}})\citenamefont {Bigi}, \citenamefont {Pegolo},
  \citenamefont {Mazitov}, \citenamefont {Schmidt},\ and\ \citenamefont
  {Ceriotti}}]{bigiPushingLimitsUnconstrained2026}%
  \BibitemOpen
  \bibfield  {author} {\bibinfo {author} {\bibfnamefont {F.}~\bibnamefont
  {Bigi}}, \bibinfo {author} {\bibfnamefont {P.}~\bibnamefont {Pegolo}},
  \bibinfo {author} {\bibfnamefont {A.}~\bibnamefont {Mazitov}}, \bibinfo
  {author} {\bibfnamefont {J.}~\bibnamefont {Schmidt}},\ and\ \bibinfo {author}
  {\bibfnamefont {M.}~\bibnamefont {Ceriotti}},\ }\href
  {https://doi.org/10.48550/arXiv.2601.16195} {\bibinfo {title} {Pushing the
  limits of unconstrained machine-learned interatomic potentials}} (\bibinfo
  {year} {2026}{\natexlab{a}}),\ \Eprint {https://arxiv.org/abs/2601.16195}
  {arXiv:2601.16195 [physics]} \BibitemShut {NoStop}%
\bibitem [{\citenamefont {Sch{\"a}fer}\ \emph {et~al.}(2026)\citenamefont
  {Sch{\"a}fer}, \citenamefont {Kellner}, \citenamefont {K{\"a}stner},\ and\
  \citenamefont {Ceriotti}}]{schaferHowTrainShallow2026}%
  \BibitemOpen
  \bibfield  {author} {\bibinfo {author} {\bibfnamefont {M.}~\bibnamefont
  {Sch{\"a}fer}}, \bibinfo {author} {\bibfnamefont {M.}~\bibnamefont
  {Kellner}}, \bibinfo {author} {\bibfnamefont {J.}~\bibnamefont
  {K{\"a}stner}},\ and\ \bibinfo {author} {\bibfnamefont {M.}~\bibnamefont
  {Ceriotti}},\ }\href {https://doi.org/10.48550/arXiv.2602.15747} {\bibinfo
  {title} {How to {{Train}} a {{Shallow Ensemble}}}} (\bibinfo {year} {2026}),\
  \Eprint {https://arxiv.org/abs/2602.15747} {arXiv:2602.15747 [physics]}
  \BibitemShut {NoStop}%
\bibitem [{\citenamefont {Torrie}\ and\ \citenamefont
  {Valleau}(1977)}]{torr-vall99jcp}%
  \BibitemOpen
  \bibfield  {author} {\bibinfo {author} {\bibfnamefont {G.~M.}\ \bibnamefont
  {Torrie}}\ and\ \bibinfo {author} {\bibfnamefont {J.~P.}\ \bibnamefont
  {Valleau}},\ }\bibfield  {title} {\bibinfo {title} {Nonphysical sampling
  distributions in {{Monte Carlo}} free-energy estimation: {{Umbrella}}
  sampling},\ }\href {https://doi.org/10.1016/0021-9991(77)90121-8} {\bibfield
  {journal} {\bibinfo  {journal} {Journal of Computational Physics}\ }\textbf
  {\bibinfo {volume} {23}},\ \bibinfo {pages} {187} (\bibinfo {year}
  {1977})}\BibitemShut {NoStop}%
\bibitem [{\citenamefont {Ceriotti}\ \emph {et~al.}(2012)\citenamefont
  {Ceriotti}, \citenamefont {Brain}, \citenamefont {Riordan},\ and\
  \citenamefont {Manolopoulos}}]{ceri+12prsa}%
  \BibitemOpen
  \bibfield  {author} {\bibinfo {author} {\bibfnamefont {M.}~\bibnamefont
  {Ceriotti}}, \bibinfo {author} {\bibfnamefont {G.~A.}\ \bibnamefont {Brain}},
  \bibinfo {author} {\bibfnamefont {O.}~\bibnamefont {Riordan}},\ and\ \bibinfo
  {author} {\bibfnamefont {D.~E.}\ \bibnamefont {Manolopoulos}},\ }\bibfield
  {title} {\bibinfo {title} {The inefficiency of re-weighted sampling and the
  curse of system size in high-order path integration},\ }\href
  {https://doi.org/10.1098/rspa.2011.0413} {\bibfield  {journal} {\bibinfo
  {journal} {Proceedings of the Royal Society A: Mathematical, Physical and
  Engineering Sciences}\ }\textbf {\bibinfo {volume} {468}},\ \bibinfo {pages}
  {2} (\bibinfo {year} {2012})},\ \Eprint {https://arxiv.org/abs/1107.1908}
  {arXiv:1107.1908} \BibitemShut {NoStop}%
\bibitem [{\citenamefont {Perdew}\ \emph {et~al.}(1996)\citenamefont {Perdew},
  \citenamefont {Burke},\ and\ \citenamefont
  {Ernzerhof}}]{perdewGeneralizedGradientApproximation1996b}%
  \BibitemOpen
  \bibfield  {author} {\bibinfo {author} {\bibfnamefont {J.~P.}\ \bibnamefont
  {Perdew}}, \bibinfo {author} {\bibfnamefont {K.}~\bibnamefont {Burke}},\ and\
  \bibinfo {author} {\bibfnamefont {M.}~\bibnamefont {Ernzerhof}},\ }\bibfield
  {title} {\bibinfo {title} {Generalized {{Gradient Approximation Made
  Simple}}},\ }\href {https://doi.org/10.1103/PhysRevLett.77.3865} {\bibfield
  {journal} {\bibinfo  {journal} {Physical Review Letters}\ }\textbf {\bibinfo
  {volume} {77}},\ \bibinfo {pages} {3865} (\bibinfo {year}
  {1996})}\BibitemShut {NoStop}%
\bibitem [{\citenamefont {Kaplan}\ \emph {et~al.}(2025)\citenamefont {Kaplan},
  \citenamefont {Liu}, \citenamefont {Qi}, \citenamefont {Ko}, \citenamefont
  {Deng}, \citenamefont {Riebesell}, \citenamefont {Ceder}, \citenamefont
  {Persson},\ and\ \citenamefont
  {Ong}}]{kaplanFoundationalPotentialEnergy2025a}%
  \BibitemOpen
  \bibfield  {author} {\bibinfo {author} {\bibfnamefont {A.~D.}\ \bibnamefont
  {Kaplan}}, \bibinfo {author} {\bibfnamefont {R.}~\bibnamefont {Liu}},
  \bibinfo {author} {\bibfnamefont {J.}~\bibnamefont {Qi}}, \bibinfo {author}
  {\bibfnamefont {T.~W.}\ \bibnamefont {Ko}}, \bibinfo {author} {\bibfnamefont
  {B.}~\bibnamefont {Deng}}, \bibinfo {author} {\bibfnamefont {J.}~\bibnamefont
  {Riebesell}}, \bibinfo {author} {\bibfnamefont {G.}~\bibnamefont {Ceder}},
  \bibinfo {author} {\bibfnamefont {K.~A.}\ \bibnamefont {Persson}},\ and\
  \bibinfo {author} {\bibfnamefont {S.~P.}\ \bibnamefont {Ong}},\ }\href
  {https://doi.org/10.48550/arXiv.2503.04070} {\bibinfo {title} {A
  {{Foundational Potential Energy Surface Dataset}} for {{Materials}}}}
  (\bibinfo {year} {2025}),\ \Eprint {https://arxiv.org/abs/2503.04070}
  {arXiv:2503.04070 [cond-mat]} \BibitemShut {NoStop}%
\bibitem [{\citenamefont {Perdew}\ \emph {et~al.}(2008)\citenamefont {Perdew},
  \citenamefont {Ruzsinszky}, \citenamefont {Csonka}, \citenamefont {Vydrov},
  \citenamefont {Scuseria}, \citenamefont {Constantin}, \citenamefont {Zhou},\
  and\ \citenamefont {Burke}}]{perdewRestoringDensityGradientExpansion2008a}%
  \BibitemOpen
  \bibfield  {author} {\bibinfo {author} {\bibfnamefont {J.~P.}\ \bibnamefont
  {Perdew}}, \bibinfo {author} {\bibfnamefont {A.}~\bibnamefont {Ruzsinszky}},
  \bibinfo {author} {\bibfnamefont {G.~I.}\ \bibnamefont {Csonka}}, \bibinfo
  {author} {\bibfnamefont {O.~A.}\ \bibnamefont {Vydrov}}, \bibinfo {author}
  {\bibfnamefont {G.~E.}\ \bibnamefont {Scuseria}}, \bibinfo {author}
  {\bibfnamefont {L.~A.}\ \bibnamefont {Constantin}}, \bibinfo {author}
  {\bibfnamefont {X.}~\bibnamefont {Zhou}},\ and\ \bibinfo {author}
  {\bibfnamefont {K.}~\bibnamefont {Burke}},\ }\bibfield  {title} {\bibinfo
  {title} {Restoring the {{Density-Gradient Expansion}} for {{Exchange}} in
  {{Solids}} and {{Surfaces}}},\ }\href
  {https://doi.org/10.1103/PhysRevLett.100.136406} {\bibfield  {journal}
  {\bibinfo  {journal} {Physical Review Letters}\ }\textbf {\bibinfo {volume}
  {100}},\ \bibinfo {pages} {136406} (\bibinfo {year} {2008})}\BibitemShut
  {NoStop}%
\bibitem [{\citenamefont {Mazitov}\ \emph
  {et~al.}(2025{\natexlab{b}})\citenamefont {Mazitov}, \citenamefont {Chorna},
  \citenamefont {Fraux}, \citenamefont {Bercx}, \citenamefont {Pizzi},
  \citenamefont {De},\ and\ \citenamefont {Ceriotti}}]{mazi+25sd}%
  \BibitemOpen
  \bibfield  {author} {\bibinfo {author} {\bibfnamefont {A.}~\bibnamefont
  {Mazitov}}, \bibinfo {author} {\bibfnamefont {S.}~\bibnamefont {Chorna}},
  \bibinfo {author} {\bibfnamefont {G.}~\bibnamefont {Fraux}}, \bibinfo
  {author} {\bibfnamefont {M.}~\bibnamefont {Bercx}}, \bibinfo {author}
  {\bibfnamefont {G.}~\bibnamefont {Pizzi}}, \bibinfo {author} {\bibfnamefont
  {S.}~\bibnamefont {De}},\ and\ \bibinfo {author} {\bibfnamefont
  {M.}~\bibnamefont {Ceriotti}},\ }\bibfield  {title} {\bibinfo {title}
  {Massive {{Atomic Diversity}}: A compact universal dataset for atomistic
  machine learning},\ }\href {https://doi.org/10.1038/s41597-025-06109-y}
  {\bibfield  {journal} {\bibinfo  {journal} {Sci Data}\ }\textbf {\bibinfo
  {volume} {12}},\ \bibinfo {pages} {1857} (\bibinfo {year}
  {2025}{\natexlab{b}})}\BibitemShut {NoStop}%
\bibitem [{\citenamefont {{Barroso-Luque}}\ \emph {et~al.}(2024)\citenamefont
  {{Barroso-Luque}}, \citenamefont {Shuaibi}, \citenamefont {Fu}, \citenamefont
  {Wood}, \citenamefont {Dzamba}, \citenamefont {Gao}, \citenamefont {Rizvi},
  \citenamefont {Zitnick},\ and\ \citenamefont
  {Ulissi}}]{barroso-luqueOpenMaterials20242024}%
  \BibitemOpen
  \bibfield  {author} {\bibinfo {author} {\bibfnamefont {L.}~\bibnamefont
  {{Barroso-Luque}}}, \bibinfo {author} {\bibfnamefont {M.}~\bibnamefont
  {Shuaibi}}, \bibinfo {author} {\bibfnamefont {X.}~\bibnamefont {Fu}},
  \bibinfo {author} {\bibfnamefont {B.~M.}\ \bibnamefont {Wood}}, \bibinfo
  {author} {\bibfnamefont {M.}~\bibnamefont {Dzamba}}, \bibinfo {author}
  {\bibfnamefont {M.}~\bibnamefont {Gao}}, \bibinfo {author} {\bibfnamefont
  {A.}~\bibnamefont {Rizvi}}, \bibinfo {author} {\bibfnamefont {C.~L.}\
  \bibnamefont {Zitnick}},\ and\ \bibinfo {author} {\bibfnamefont {Z.~W.}\
  \bibnamefont {Ulissi}},\ }\href {https://doi.org/10.48550/arXiv.2410.12771}
  {\bibinfo {title} {Open {{Materials}} 2024 ({{OMat24}}) {{Inorganic Materials
  Dataset}} and {{Models}}}} (\bibinfo {year} {2024}),\ \Eprint
  {https://arxiv.org/abs/2410.12771} {arXiv:2410.12771 [cond-mat]} \BibitemShut
  {NoStop}%
\bibitem [{\citenamefont {Furness}\ \emph {et~al.}(2020)\citenamefont
  {Furness}, \citenamefont {Kaplan}, \citenamefont {Ning}, \citenamefont
  {Perdew},\ and\ \citenamefont
  {Sun}}]{furnessAccurateNumericallyEfficient2020}%
  \BibitemOpen
  \bibfield  {author} {\bibinfo {author} {\bibfnamefont {J.~W.}\ \bibnamefont
  {Furness}}, \bibinfo {author} {\bibfnamefont {A.~D.}\ \bibnamefont {Kaplan}},
  \bibinfo {author} {\bibfnamefont {J.}~\bibnamefont {Ning}}, \bibinfo {author}
  {\bibfnamefont {J.~P.}\ \bibnamefont {Perdew}},\ and\ \bibinfo {author}
  {\bibfnamefont {J.}~\bibnamefont {Sun}},\ }\bibfield  {title} {\bibinfo
  {title} {Accurate and {{Numerically Efficient r2SCAN Meta-Generalized
  Gradient Approximation}}},\ }\href
  {https://doi.org/10.1021/acs.jpclett.0c02405} {\bibfield  {journal} {\bibinfo
   {journal} {The Journal of Physical Chemistry Letters}\ }\textbf {\bibinfo
  {volume} {11}},\ \bibinfo {pages} {8208} (\bibinfo {year}
  {2020})}\BibitemShut {NoStop}%
\bibitem [{\citenamefont {Grimme}\ \emph {et~al.}(2010)\citenamefont {Grimme},
  \citenamefont {Antony}, \citenamefont {Ehrlich},\ and\ \citenamefont
  {Krieg}}]{grimmeConsistentAccurateInitio2010a}%
  \BibitemOpen
  \bibfield  {author} {\bibinfo {author} {\bibfnamefont {S.}~\bibnamefont
  {Grimme}}, \bibinfo {author} {\bibfnamefont {J.}~\bibnamefont {Antony}},
  \bibinfo {author} {\bibfnamefont {S.}~\bibnamefont {Ehrlich}},\ and\ \bibinfo
  {author} {\bibfnamefont {H.}~\bibnamefont {Krieg}},\ }\bibfield  {title}
  {\bibinfo {title} {A consistent and accurate ab initio parametrization of
  density functional dispersion correction ({{DFT-D}}) for the 94 elements
  {{H-Pu}}},\ }\href {https://doi.org/10.1063/1.3382344} {\bibfield  {journal}
  {\bibinfo  {journal} {The Journal of Chemical Physics}\ }\textbf {\bibinfo
  {volume} {132}},\ \bibinfo {pages} {154104} (\bibinfo {year}
  {2010})}\BibitemShut {NoStop}%
\bibitem [{\citenamefont {Curtiss}\ \emph {et~al.}(2000)\citenamefont
  {Curtiss}, \citenamefont {Raghavachari}, \citenamefont {Redfern},\ and\
  \citenamefont {Pople}}]{curtissAssessmentGaussian3Density2000}%
  \BibitemOpen
  \bibfield  {author} {\bibinfo {author} {\bibfnamefont {L.~A.}\ \bibnamefont
  {Curtiss}}, \bibinfo {author} {\bibfnamefont {K.}~\bibnamefont
  {Raghavachari}}, \bibinfo {author} {\bibfnamefont {P.~C.}\ \bibnamefont
  {Redfern}},\ and\ \bibinfo {author} {\bibfnamefont {J.~A.}\ \bibnamefont
  {Pople}},\ }\bibfield  {title} {\bibinfo {title} {Assessment of
  {{Gaussian-3}} and density functional theories for a larger experimental test
  set},\ }\href {https://doi.org/10.1063/1.481336} {\bibfield  {journal}
  {\bibinfo  {journal} {The Journal of Chemical Physics}\ }\textbf {\bibinfo
  {volume} {112}},\ \bibinfo {pages} {7374} (\bibinfo {year}
  {2000})}\BibitemShut {NoStop}%
\bibitem [{\citenamefont {Tran}\ \emph {et~al.}(2016)\citenamefont {Tran},
  \citenamefont {Stelzl},\ and\ \citenamefont {Blaha}}]{tranRungs142016}%
  \BibitemOpen
  \bibfield  {author} {\bibinfo {author} {\bibfnamefont {F.}~\bibnamefont
  {Tran}}, \bibinfo {author} {\bibfnamefont {J.}~\bibnamefont {Stelzl}},\ and\
  \bibinfo {author} {\bibfnamefont {P.}~\bibnamefont {Blaha}},\ }\bibfield
  {title} {\bibinfo {title} {Rungs 1 to 4 of {{DFT Jacob}}'s ladder:
  {{Extensive}} test on the lattice constant, bulk modulus, and cohesive energy
  of solids},\ }\href {https://doi.org/10.1063/1.4948636} {\bibfield  {journal}
  {\bibinfo  {journal} {The Journal of Chemical Physics}\ }\textbf {\bibinfo
  {volume} {144}},\ \bibinfo {pages} {204120} (\bibinfo {year}
  {2016})}\BibitemShut {NoStop}%
\bibitem [{\citenamefont {Bussi}\ \emph {et~al.}(2007)\citenamefont {Bussi},
  \citenamefont {Donadio},\ and\ \citenamefont {Parrinello}}]{buss+07jcp}%
  \BibitemOpen
  \bibfield  {author} {\bibinfo {author} {\bibfnamefont {G.}~\bibnamefont
  {Bussi}}, \bibinfo {author} {\bibfnamefont {D.}~\bibnamefont {Donadio}},\
  and\ \bibinfo {author} {\bibfnamefont {M.}~\bibnamefont {Parrinello}},\
  }\bibfield  {title} {\bibinfo {title} {Canonical sampling through velocity
  rescaling},\ }\href@noop {} {\bibfield  {journal} {\bibinfo  {journal}
  {Journal of Chemical Physics}\ }\textbf {\bibinfo {volume} {126}},\ \bibinfo
  {pages} {14101} (\bibinfo {year} {2007})}\BibitemShut {NoStop}%
\bibitem [{\citenamefont {Ceriotti}\ \emph {et~al.}(2010)\citenamefont
  {Ceriotti}, \citenamefont {Bussi},\ and\ \citenamefont
  {Parrinello}}]{ceri+10jctc}%
  \BibitemOpen
  \bibfield  {author} {\bibinfo {author} {\bibfnamefont {M.}~\bibnamefont
  {Ceriotti}}, \bibinfo {author} {\bibfnamefont {G.}~\bibnamefont {Bussi}},\
  and\ \bibinfo {author} {\bibfnamefont {M.}~\bibnamefont {Parrinello}},\
  }\bibfield  {title} {\bibinfo {title} {Colored-noise thermostats \`a la
  {{Carte}}},\ }\href {https://doi.org/10.1021/ct900563s} {\bibfield  {journal}
  {\bibinfo  {journal} {Journal of Chemical Theory and Computation}\ }\textbf
  {\bibinfo {volume} {6}},\ \bibinfo {pages} {1170} (\bibinfo {year}
  {2010})}\BibitemShut {NoStop}%
\bibitem [{\citenamefont {Bussi}\ \emph {et~al.}(2009)\citenamefont {Bussi},
  \citenamefont {{Zykova-Timan}},\ and\ \citenamefont
  {Parrinello}}]{buss+09jcp}%
  \BibitemOpen
  \bibfield  {author} {\bibinfo {author} {\bibfnamefont {G.}~\bibnamefont
  {Bussi}}, \bibinfo {author} {\bibfnamefont {T.}~\bibnamefont
  {{Zykova-Timan}}},\ and\ \bibinfo {author} {\bibfnamefont {M.}~\bibnamefont
  {Parrinello}},\ }\bibfield  {title} {\bibinfo {title} {Isothermal-isobaric
  molecular dynamics using stochastic velocity rescaling},\ }\href
  {https://doi.org/10.1063/1.3073889} {\bibfield  {journal} {\bibinfo
  {journal} {The Journal of Chemical Physics}\ }\textbf {\bibinfo {volume}
  {130}},\ \bibinfo {pages} {074101} (\bibinfo {year} {2009})}\BibitemShut
  {NoStop}%
\bibitem [{\citenamefont {Ceriotti}\ \emph {et~al.}(2014)\citenamefont
  {Ceriotti}, \citenamefont {More},\ and\ \citenamefont
  {Manolopoulos}}]{ceri+14cpc}%
  \BibitemOpen
  \bibfield  {author} {\bibinfo {author} {\bibfnamefont {M.}~\bibnamefont
  {Ceriotti}}, \bibinfo {author} {\bibfnamefont {J.}~\bibnamefont {More}},\
  and\ \bibinfo {author} {\bibfnamefont {D.~E.}\ \bibnamefont {Manolopoulos}},\
  }\bibfield  {title} {\bibinfo {title} {I-{{PI}}: {{A Python}} interface for
  ab initio path integral molecular dynamics simulations},\ }\href
  {https://doi.org/10.1016/j.cpc.2013.10.027} {\bibfield  {journal} {\bibinfo
  {journal} {Computer Physics Communications}\ }\textbf {\bibinfo {volume}
  {185}},\ \bibinfo {pages} {1019} (\bibinfo {year} {2014})}\BibitemShut
  {NoStop}%
\bibitem [{\citenamefont {Litman}\ \emph {et~al.}(2024)\citenamefont {Litman},
  \citenamefont {Kapil}, \citenamefont {Feldman}, \citenamefont {Tisi},
  \citenamefont {Begu{\v s}i{\'c}}, \citenamefont {Fidanyan}, \citenamefont
  {Fraux}, \citenamefont {Higer}, \citenamefont {Kellner}, \citenamefont {Li},
  \citenamefont {P{\'o}s}, \citenamefont {Stocco}, \citenamefont {Trenins},
  \citenamefont {Hirshberg}, \citenamefont {Rossi},\ and\ \citenamefont
  {Ceriotti}}]{litm+24jcp}%
  \BibitemOpen
  \bibfield  {author} {\bibinfo {author} {\bibfnamefont {Y.}~\bibnamefont
  {Litman}}, \bibinfo {author} {\bibfnamefont {V.}~\bibnamefont {Kapil}},
  \bibinfo {author} {\bibfnamefont {Y.~M.~Y.}\ \bibnamefont {Feldman}},
  \bibinfo {author} {\bibfnamefont {D.}~\bibnamefont {Tisi}}, \bibinfo {author}
  {\bibfnamefont {T.}~\bibnamefont {Begu{\v s}i{\'c}}}, \bibinfo {author}
  {\bibfnamefont {K.}~\bibnamefont {Fidanyan}}, \bibinfo {author}
  {\bibfnamefont {G.}~\bibnamefont {Fraux}}, \bibinfo {author} {\bibfnamefont
  {J.}~\bibnamefont {Higer}}, \bibinfo {author} {\bibfnamefont
  {M.}~\bibnamefont {Kellner}}, \bibinfo {author} {\bibfnamefont {T.~E.}\
  \bibnamefont {Li}}, \bibinfo {author} {\bibfnamefont {E.~S.}\ \bibnamefont
  {P{\'o}s}}, \bibinfo {author} {\bibfnamefont {E.}~\bibnamefont {Stocco}},
  \bibinfo {author} {\bibfnamefont {G.}~\bibnamefont {Trenins}}, \bibinfo
  {author} {\bibfnamefont {B.}~\bibnamefont {Hirshberg}}, \bibinfo {author}
  {\bibfnamefont {M.}~\bibnamefont {Rossi}},\ and\ \bibinfo {author}
  {\bibfnamefont {M.}~\bibnamefont {Ceriotti}},\ }\bibfield  {title} {\bibinfo
  {title} {I-{{PI}} 3.0: {{A}} flexible and efficient framework for advanced
  atomistic simulations},\ }\href {https://doi.org/10.1063/5.0215869}
  {\bibfield  {journal} {\bibinfo  {journal} {The Journal of Chemical Physics}\
  }\textbf {\bibinfo {volume} {161}},\ \bibinfo {pages} {062504} (\bibinfo
  {year} {2024})}\BibitemShut {NoStop}%
\bibitem [{\citenamefont {Bigi}\ \emph
  {et~al.}(2026{\natexlab{b}})\citenamefont {Bigi}, \citenamefont {Abbott},
  \citenamefont {Loche}, \citenamefont {Mazitov}, \citenamefont {Tisi},
  \citenamefont {Langer}, \citenamefont {Goscinski}, \citenamefont {Pegolo},
  \citenamefont {Chong}, \citenamefont {Goswami}, \citenamefont {Febrer},
  \citenamefont {Chorna}, \citenamefont {Kellner}, \citenamefont {Ceriotti},\
  and\ \citenamefont {Fraux}}]{bigi+26jcp}%
  \BibitemOpen
  \bibfield  {author} {\bibinfo {author} {\bibfnamefont {F.}~\bibnamefont
  {Bigi}}, \bibinfo {author} {\bibfnamefont {J.~W.}\ \bibnamefont {Abbott}},
  \bibinfo {author} {\bibfnamefont {P.}~\bibnamefont {Loche}}, \bibinfo
  {author} {\bibfnamefont {A.}~\bibnamefont {Mazitov}}, \bibinfo {author}
  {\bibfnamefont {D.}~\bibnamefont {Tisi}}, \bibinfo {author} {\bibfnamefont
  {M.~F.}\ \bibnamefont {Langer}}, \bibinfo {author} {\bibfnamefont
  {A.}~\bibnamefont {Goscinski}}, \bibinfo {author} {\bibfnamefont
  {P.}~\bibnamefont {Pegolo}}, \bibinfo {author} {\bibfnamefont
  {S.}~\bibnamefont {Chong}}, \bibinfo {author} {\bibfnamefont
  {R.}~\bibnamefont {Goswami}}, \bibinfo {author} {\bibfnamefont
  {P.}~\bibnamefont {Febrer}}, \bibinfo {author} {\bibfnamefont
  {S.}~\bibnamefont {Chorna}}, \bibinfo {author} {\bibfnamefont
  {M.}~\bibnamefont {Kellner}}, \bibinfo {author} {\bibfnamefont
  {M.}~\bibnamefont {Ceriotti}},\ and\ \bibinfo {author} {\bibfnamefont
  {G.}~\bibnamefont {Fraux}},\ }\bibfield  {title} {\bibinfo {title}
  {Metatensor and metatomic : {{Foundational}} libraries for interoperable
  atomistic machine learning},\ }\href {https://doi.org/10.1063/5.0304911}
  {\bibfield  {journal} {\bibinfo  {journal} {The Journal of Chemical Physics}\
  }\textbf {\bibinfo {volume} {164}},\ \bibinfo {pages} {064113} (\bibinfo
  {year} {2026}{\natexlab{b}})}\BibitemShut {NoStop}%
\bibitem [{\citenamefont {Gowers}\ \emph {et~al.}(2016)\citenamefont {Gowers},
  \citenamefont {Linke}, \citenamefont {Barnoud}, \citenamefont {Reddy},
  \citenamefont {Melo}, \citenamefont {Seyler}, \citenamefont {Doma{\'n}ski},
  \citenamefont {Dotson}, \citenamefont {Buchoux}, \citenamefont {Kenney},\
  and\ \citenamefont {Beckstein}}]{gowersMDAnalysisPythonPackage2016}%
  \BibitemOpen
  \bibfield  {author} {\bibinfo {author} {\bibfnamefont {R.~J.}\ \bibnamefont
  {Gowers}}, \bibinfo {author} {\bibfnamefont {M.}~\bibnamefont {Linke}},
  \bibinfo {author} {\bibfnamefont {J.}~\bibnamefont {Barnoud}}, \bibinfo
  {author} {\bibfnamefont {T.~J.~E.}\ \bibnamefont {Reddy}}, \bibinfo {author}
  {\bibfnamefont {M.~N.}\ \bibnamefont {Melo}}, \bibinfo {author}
  {\bibfnamefont {S.~L.}\ \bibnamefont {Seyler}}, \bibinfo {author}
  {\bibfnamefont {J.}~\bibnamefont {Doma{\'n}ski}}, \bibinfo {author}
  {\bibfnamefont {D.~L.}\ \bibnamefont {Dotson}}, \bibinfo {author}
  {\bibfnamefont {S.}~\bibnamefont {Buchoux}}, \bibinfo {author} {\bibfnamefont
  {I.~M.}\ \bibnamefont {Kenney}},\ and\ \bibinfo {author} {\bibfnamefont
  {O.}~\bibnamefont {Beckstein}},\ }\bibfield  {title} {\bibinfo {title}
  {{{MDAnalysis}}: {{A Python Package}} for the {{Rapid Analysis}} of
  {{Molecular Dynamics Simulations}}},\ }\bibfield  {journal} {\bibinfo
  {journal} {SciPy 2016}\ }\href {https://doi.org/10.25080/Majora-629e541a-00e}
  {10.25080/Majora-629e541a-00e} (\bibinfo {year} {2016})\BibitemShut {NoStop}%
\bibitem [{\citenamefont {{Michaud-Agrawal}}\ \emph {et~al.}(2011)\citenamefont
  {{Michaud-Agrawal}}, \citenamefont {Denning}, \citenamefont {Woolf},\ and\
  \citenamefont {Beckstein}}]{michaud-agrawalMDAnalysisToolkitAnalysis2011}%
  \BibitemOpen
  \bibfield  {author} {\bibinfo {author} {\bibfnamefont {N.}~\bibnamefont
  {{Michaud-Agrawal}}}, \bibinfo {author} {\bibfnamefont {E.~J.}\ \bibnamefont
  {Denning}}, \bibinfo {author} {\bibfnamefont {T.~B.}\ \bibnamefont {Woolf}},\
  and\ \bibinfo {author} {\bibfnamefont {O.}~\bibnamefont {Beckstein}},\
  }\bibfield  {title} {\bibinfo {title} {{{MDAnalysis}}: {{A}} toolkit for the
  analysis of molecular dynamics simulations},\ }\href
  {https://doi.org/10.1002/jcc.21787} {\bibfield  {journal} {\bibinfo
  {journal} {Journal of Computational Chemistry}\ }\textbf {\bibinfo {volume}
  {32}},\ \bibinfo {pages} {2319} (\bibinfo {year} {2011})}\BibitemShut
  {NoStop}%
\bibitem [{\citenamefont {Petkov}(2012)}]{petkovPairDistributionFunctions2012}%
  \BibitemOpen
  \bibfield  {author} {\bibinfo {author} {\bibfnamefont {V.}~\bibnamefont
  {Petkov}},\ }\bibfield  {title} {\bibinfo {title} {Pair {{Distribution
  Functions Analysis}}},\ }in\ \href
  {https://doi.org/10.1002/0471266965.com159} {\emph {\bibinfo {booktitle}
  {Characterization of {{Materials}}}}}\ (\bibinfo  {publisher} {John Wiley \&
  Sons, Ltd},\ \bibinfo {year} {2012})\ pp.\ \bibinfo {pages}
  {1--14}\BibitemShut {NoStop}%
\bibitem [{\citenamefont {Tovey}\ \emph {et~al.}(2020)\citenamefont {Tovey},
  \citenamefont {Narayanan~Krishnamoorthy}, \citenamefont {Sivaraman},
  \citenamefont {Guo}, \citenamefont {Benmore}, \citenamefont {Heuer},\ and\
  \citenamefont {Holm}}]{toveyDFTAccurateInteratomic2020}%
  \BibitemOpen
  \bibfield  {author} {\bibinfo {author} {\bibfnamefont {S.}~\bibnamefont
  {Tovey}}, \bibinfo {author} {\bibfnamefont {A.}~\bibnamefont
  {Narayanan~Krishnamoorthy}}, \bibinfo {author} {\bibfnamefont
  {G.}~\bibnamefont {Sivaraman}}, \bibinfo {author} {\bibfnamefont
  {J.}~\bibnamefont {Guo}}, \bibinfo {author} {\bibfnamefont {C.}~\bibnamefont
  {Benmore}}, \bibinfo {author} {\bibfnamefont {A.}~\bibnamefont {Heuer}},\
  and\ \bibinfo {author} {\bibfnamefont {C.}~\bibnamefont {Holm}},\ }\bibfield
  {title} {\bibinfo {title} {{{DFT Accurate Interatomic Potential}} for
  {{Molten NaCl}} from {{Machine Learning}}},\ }\href
  {https://doi.org/10.1021/acs.jpcc.0c08870} {\bibfield  {journal} {\bibinfo
  {journal} {The Journal of Physical Chemistry C}\ }\textbf {\bibinfo {volume}
  {124}},\ \bibinfo {pages} {25760} (\bibinfo {year} {2020})}\BibitemShut
  {NoStop}%
\bibitem [{\citenamefont {Ohno}\ and\ \citenamefont
  {Furukawa}(1981)}]{ohnoXrayDiffractionAnalysis1981}%
  \BibitemOpen
  \bibfield  {author} {\bibinfo {author} {\bibfnamefont {H.}~\bibnamefont
  {Ohno}}\ and\ \bibinfo {author} {\bibfnamefont {K.}~\bibnamefont
  {Furukawa}},\ }\bibfield  {title} {\bibinfo {title} {X-ray diffraction
  analysis of molten {{NaCl}} near its melting point},\ }\href
  {https://doi.org/10.1039/F19817701981} {\bibfield  {journal} {\bibinfo
  {journal} {Journal of the Chemical Society, Faraday Transactions 1: Physical
  Chemistry in Condensed Phases}\ }\textbf {\bibinfo {volume} {77}},\ \bibinfo
  {pages} {1981} (\bibinfo {year} {1981})}\BibitemShut {NoStop}%
\bibitem [{\citenamefont {{Del Rio}}\ \emph {et~al.}(2020)\citenamefont {{Del
  Rio}}, \citenamefont {{Pascual}}, \citenamefont {{Rodriguez}}, \citenamefont
  {{Gonz{\'a}lez}},\ and\ \citenamefont
  {{Gonz{\'a}lez}}}]{delrioFirstPrinciplesDetermination2020}%
  \BibitemOpen
  \bibfield  {author} {\bibinfo {author} {\bibnamefont {{Del Rio}}}, \bibinfo
  {author} {\bibnamefont {{Pascual}}}, \bibinfo {author} {\bibnamefont
  {{Rodriguez}}}, \bibinfo {author} {\bibnamefont {{Gonz{\'a}lez}}},\ and\
  \bibinfo {author} {\bibnamefont {{Gonz{\'a}lez}}},\ }\bibfield  {title}
  {\bibinfo {title} {First principles determination of some static and dynamic
  properties of the liquid 3d transition metals near melting},\ }\href
  {https://doi.org/10.5488/CMP.23.23606} {\bibfield  {journal} {\bibinfo
  {journal} {Condensed Matter Physics}\ }\textbf {\bibinfo {volume} {23}},\
  \bibinfo {pages} {23606} (\bibinfo {year} {2020})}\BibitemShut {NoStop}%
\bibitem [{\citenamefont
  {Soper}(2000)}]{soperRadialDistributionFunctions2000c}%
  \BibitemOpen
  \bibfield  {author} {\bibinfo {author} {\bibfnamefont {A.~K.}\ \bibnamefont
  {Soper}},\ }\bibfield  {title} {\bibinfo {title} {The radial distribution
  functions of water and ice from 220 to 673 {{K}} and at pressures up to 400
  {{MPa}}},\ }\href {https://doi.org/10.1016/S0301-0104(00)00179-8} {\bibfield
  {journal} {\bibinfo  {journal} {Chemical Physics}\ }\textbf {\bibinfo
  {volume} {258}},\ \bibinfo {pages} {121} (\bibinfo {year}
  {2000})}\BibitemShut {NoStop}%
\bibitem [{\citenamefont {Lide}(2005)}]{CRCDensityWater2005}%
  \BibitemOpen
  \bibinfo {editor} {\bibfnamefont {D.~R.}\ \bibnamefont {Lide}},\ ed.,\
  \bibinfo {title} {Density of water at 1 atmosphere},\ in\ \href
  {http://www.hbcpnetbase.com} {\emph {\bibinfo {booktitle} {CRC Handbook of
  Chemistry and Physics}}}\ (\bibinfo  {publisher} {CRC Press},\ \bibinfo
  {address} {Boca Raton, FL},\ \bibinfo {year} {2005})\ Chap.~\bibinfo
  {chapter} {6}, pp.\ \bibinfo {pages} {{6--5}},\ \bibinfo {edition} {internet
  version 2005}\ ed.\BibitemShut {Stop}%
\bibitem [{\citenamefont {Bellissent}\ \emph {et~al.}(1987)\citenamefont
  {Bellissent}, \citenamefont {Bergman}, \citenamefont {Ceolin},\ and\
  \citenamefont {Gaspard}}]{bellissentStructureLiquidPeierls1987}%
  \BibitemOpen
  \bibfield  {author} {\bibinfo {author} {\bibfnamefont {R.}~\bibnamefont
  {Bellissent}}, \bibinfo {author} {\bibfnamefont {C.}~\bibnamefont {Bergman}},
  \bibinfo {author} {\bibfnamefont {R.}~\bibnamefont {Ceolin}},\ and\ \bibinfo
  {author} {\bibfnamefont {J.~P.}\ \bibnamefont {Gaspard}},\ }\bibfield
  {title} {\bibinfo {title} {Structure of liquid {{As}}: {{A Peierls}}
  distortion in a liquid},\ }\href {https://doi.org/10.1103/PhysRevLett.59.661}
  {\bibfield  {journal} {\bibinfo  {journal} {Physical Review Letters}\
  }\textbf {\bibinfo {volume} {59}},\ \bibinfo {pages} {661} (\bibinfo {year}
  {1987})}\BibitemShut {NoStop}%
\bibitem [{\citenamefont {Klemm}\ and\ \citenamefont
  {Niermann}(1963)}]{klemmFurtherContributionsKnowledge1963}%
  \BibitemOpen
  \bibfield  {author} {\bibinfo {author} {\bibfnamefont {W.}~\bibnamefont
  {Klemm}}\ and\ \bibinfo {author} {\bibfnamefont {H.}~\bibnamefont
  {Niermann}},\ }\bibfield  {title} {\bibinfo {title} {Further
  {{Contributions}} to the {{Knowledge}} of {{Semimetals}}},\ }\href
  {https://doi.org/10.1002/anie.196305231} {\bibfield  {journal} {\bibinfo
  {journal} {Angewandte Chemie International Edition in English}\ }\textbf
  {\bibinfo {volume} {2}},\ \bibinfo {pages} {523} (\bibinfo {year}
  {1963})}\BibitemShut {NoStop}%
\bibitem [{\citenamefont {{Bellissent-Funel}}\ \emph
  {et~al.}(1989)\citenamefont {{Bellissent-Funel}}, \citenamefont {Chieux},
  \citenamefont {Levesque},\ and\ \citenamefont
  {Weis}}]{bellissent-funelStructureFactorEffective1989}%
  \BibitemOpen
  \bibfield  {author} {\bibinfo {author} {\bibfnamefont {M.~C.}\ \bibnamefont
  {{Bellissent-Funel}}}, \bibinfo {author} {\bibfnamefont {P.}~\bibnamefont
  {Chieux}}, \bibinfo {author} {\bibfnamefont {D.}~\bibnamefont {Levesque}},\
  and\ \bibinfo {author} {\bibfnamefont {J.~J.}\ \bibnamefont {Weis}},\
  }\bibfield  {title} {\bibinfo {title} {Structure factor and effective
  two-body potential for liquid gallium},\ }\href
  {https://doi.org/10.1103/PhysRevA.39.6310} {\bibfield  {journal} {\bibinfo
  {journal} {Physical Review A}\ }\textbf {\bibinfo {volume} {39}},\ \bibinfo
  {pages} {6310} (\bibinfo {year} {1989})}\BibitemShut {NoStop}%
\bibitem [{\citenamefont {Bergman}\ \emph {et~al.}(1985)\citenamefont
  {Bergman}, \citenamefont {Bichara}, \citenamefont {Chieux},\ and\
  \citenamefont {Gaspard}}]{bergmanATOMICSTRUCTURELIQUID1985}%
  \BibitemOpen
  \bibfield  {author} {\bibinfo {author} {\bibfnamefont {C.}~\bibnamefont
  {Bergman}}, \bibinfo {author} {\bibfnamefont {C.}~\bibnamefont {Bichara}},
  \bibinfo {author} {\bibfnamefont {P.}~\bibnamefont {Chieux}},\ and\ \bibinfo
  {author} {\bibfnamefont {J.~P.}\ \bibnamefont {Gaspard}},\ }\bibfield
  {title} {\bibinfo {title} {{{ON THE ATOMIC STRUCTURE OF LIQUID GaAs}}},\
  }\href {https://doi.org/10.1051/jphyscol:1985811} {\bibfield  {journal}
  {\bibinfo  {journal} {Le Journal de Physique Colloques}\ }\textbf {\bibinfo
  {volume} {46}},\ \bibinfo {pages} {C8} (\bibinfo {year} {1985})}\BibitemShut
  {NoStop}%
\bibitem [{\citenamefont {Glazov}\ \emph {et~al.}(1969)\citenamefont {Glazov},
  \citenamefont {Chizhevskaya},\ and\ \citenamefont
  {Glagoleva}}]{Glazov1969LiquidSemiconductors}%
  \BibitemOpen
  \bibfield  {author} {\bibinfo {author} {\bibfnamefont {V.~M.}\ \bibnamefont
  {Glazov}}, \bibinfo {author} {\bibfnamefont {S.~N.}\ \bibnamefont
  {Chizhevskaya}},\ and\ \bibinfo {author} {\bibfnamefont {N.~N.}\ \bibnamefont
  {Glagoleva}},\ }\href@noop {} {\emph {\bibinfo {title} {Liquid
  Semiconductors}}}\ (\bibinfo  {publisher} {Plenum Press},\ \bibinfo {year}
  {1969})\ p.\ \bibinfo {pages} {124}\BibitemShut {NoStop}%
\bibitem [{\citenamefont {Waseda}(1980)}]{waseda1980structure}%
  \BibitemOpen
  \bibfield  {author} {\bibinfo {author} {\bibfnamefont {Y.}~\bibnamefont
  {Waseda}},\ }\bibfield  {title} {\bibinfo {title} {The structure of
  non-crystalline materials},\ }\href@noop {} {\bibfield  {journal} {\bibinfo
  {journal} {Liquids and Amorphous Solids}\ } (\bibinfo {year}
  {1980})}\BibitemShut {NoStop}%
\bibitem [{\citenamefont {Lou}\ \emph {et~al.}(2013)\citenamefont {Lou},
  \citenamefont {Wang}, \citenamefont {Cao}, \citenamefont {Zhang},
  \citenamefont {Zhang}, \citenamefont {Hu}, \citenamefont {Mao},\ and\
  \citenamefont {Jiang}}]{louNegativeExpansionsInteratomic2013}%
  \BibitemOpen
  \bibfield  {author} {\bibinfo {author} {\bibfnamefont {H.}~\bibnamefont
  {Lou}}, \bibinfo {author} {\bibfnamefont {X.}~\bibnamefont {Wang}}, \bibinfo
  {author} {\bibfnamefont {Q.}~\bibnamefont {Cao}}, \bibinfo {author}
  {\bibfnamefont {D.}~\bibnamefont {Zhang}}, \bibinfo {author} {\bibfnamefont
  {J.}~\bibnamefont {Zhang}}, \bibinfo {author} {\bibfnamefont
  {T.}~\bibnamefont {Hu}}, \bibinfo {author} {\bibfnamefont {H.-k.}\
  \bibnamefont {Mao}},\ and\ \bibinfo {author} {\bibfnamefont {J.-Z.}\
  \bibnamefont {Jiang}},\ }\bibfield  {title} {\bibinfo {title} {Negative
  expansions of interatomic distances in metallic melts},\ }\href
  {https://doi.org/10.1073/pnas.1307967110} {\bibfield  {journal} {\bibinfo
  {journal} {Proceedings of the National Academy of Sciences}\ }\textbf
  {\bibinfo {volume} {110}},\ \bibinfo {pages} {10068} (\bibinfo {year}
  {2013})}\BibitemShut {NoStop}%
\bibitem [{\citenamefont {{Schmitz-Pranghe}}\ and\ \citenamefont
  {Kohlhaas}(1970)}]{schmitz-prangheNotizenRontgenbeugungsuntersuchungenEisen1970}%
  \BibitemOpen
  \bibfield  {author} {\bibinfo {author} {\bibfnamefont {N.}~\bibnamefont
  {{Schmitz-Pranghe}}}\ and\ \bibinfo {author} {\bibfnamefont {R.}~\bibnamefont
  {Kohlhaas}},\ }\bibfield  {title} {\bibinfo {title} {Notizen:
  {{R{\"o}ntgenbeugungsuntersuchungen}} an {{Eisen}}, {{Kobalt}} und {{Nickel}}
  im fl{\"u}ssigen {{Zustand}}},\ }\href
  {https://doi.org/10.1515/zna-1970-1133} {\bibfield  {journal} {\bibinfo
  {journal} {Zeitschrift f{\"u}r Naturforschung A}\ }\textbf {\bibinfo {volume}
  {25}},\ \bibinfo {pages} {1752} (\bibinfo {year} {1970})}\BibitemShut
  {NoStop}%
\bibitem [{\citenamefont {Mendelev}\ \emph {et~al.}(2003)\citenamefont
  {Mendelev}, \citenamefont {Han}, \citenamefont {Srolovitz}, \citenamefont
  {Ackland}, \citenamefont {Sun},\ and\ \citenamefont
  {Asta}}]{mendelevDevelopmentNewInteratomic2003}%
  \BibitemOpen
  \bibfield  {author} {\bibinfo {author} {\bibfnamefont {M.~I.}\ \bibnamefont
  {Mendelev}}, \bibinfo {author} {\bibfnamefont {S.}~\bibnamefont {Han}},
  \bibinfo {author} {\bibfnamefont {D.~J.}\ \bibnamefont {Srolovitz}}, \bibinfo
  {author} {\bibfnamefont {G.~J.}\ \bibnamefont {Ackland}}, \bibinfo {author}
  {\bibfnamefont {D.~Y.}\ \bibnamefont {Sun}},\ and\ \bibinfo {author}
  {\bibfnamefont {M.}~\bibnamefont {Asta}},\ }\bibfield  {title} {\bibinfo
  {title} {Development of new interatomic potentials appropriate for
  crystalline and liquid iron},\ }\href
  {https://doi.org/10.1080/14786430310001613264} {\bibfield  {journal}
  {\bibinfo  {journal} {Philosophical Magazine}\ }\textbf {\bibinfo {volume}
  {83}},\ \bibinfo {pages} {3977} (\bibinfo {year} {2003})}\BibitemShut
  {NoStop}%
\bibitem [{\citenamefont {Il'inskii}\ \emph {et~al.}(2002)\citenamefont
  {Il'inskii}, \citenamefont {Slyusarenko}, \citenamefont {Slukhovskii},
  \citenamefont {Kaban},\ and\ \citenamefont
  {Hoyer}}]{ilinskiiStructureLiquidFe2002}%
  \BibitemOpen
  \bibfield  {author} {\bibinfo {author} {\bibfnamefont {A.}~\bibnamefont
  {Il'inskii}}, \bibinfo {author} {\bibfnamefont {S.}~\bibnamefont
  {Slyusarenko}}, \bibinfo {author} {\bibfnamefont {O.}~\bibnamefont
  {Slukhovskii}}, \bibinfo {author} {\bibfnamefont {I.}~\bibnamefont {Kaban}},\
  and\ \bibinfo {author} {\bibfnamefont {W.}~\bibnamefont {Hoyer}},\ }\bibfield
   {title} {\bibinfo {title} {Structure of liquid {{Fe}}--{{Al}} alloys},\
  }\href {https://doi.org/10.1016/S0921-5093(01)01457-5} {\bibfield  {journal}
  {\bibinfo  {journal} {Materials Science and Engineering: A}\ }\textbf
  {\bibinfo {volume} {325}},\ \bibinfo {pages} {98} (\bibinfo {year}
  {2002})}\BibitemShut {NoStop}%
\bibitem [{\citenamefont {Ntonti}\ \emph {et~al.}(2024)\citenamefont {Ntonti},
  \citenamefont {Sotiriadou}, \citenamefont {Assael}, \citenamefont {Huber},
  \citenamefont {Wilthan},\ and\ \citenamefont
  {Watanabe}}]{ntontiReferenceCorrelationsDensity2024}%
  \BibitemOpen
  \bibfield  {author} {\bibinfo {author} {\bibfnamefont {E.}~\bibnamefont
  {Ntonti}}, \bibinfo {author} {\bibfnamefont {S.}~\bibnamefont {Sotiriadou}},
  \bibinfo {author} {\bibfnamefont {M.~J.}\ \bibnamefont {Assael}}, \bibinfo
  {author} {\bibfnamefont {M.~L.}\ \bibnamefont {Huber}}, \bibinfo {author}
  {\bibfnamefont {B.}~\bibnamefont {Wilthan}},\ and\ \bibinfo {author}
  {\bibfnamefont {M.}~\bibnamefont {Watanabe}},\ }\bibfield  {title} {\bibinfo
  {title} {Reference {{Correlations}} for the {{Density}} and {{Thermal
  Conductivity}}, and {{Review}} of the {{Viscosity Measurements}}, of {{Liquid
  Titanium}}, {{Zirconium}}, {{Hafnium}}, {{Vanadium}}, {{Niobium}},
  {{Tantalum}}, {{Chromium}}, {{Molybdenum}}, and {{Tungsten}}},\ }\href
  {https://doi.org/10.1007/s10765-023-03305-z} {\bibfield  {journal} {\bibinfo
  {journal} {International Journal of Thermophysics}\ }\textbf {\bibinfo
  {volume} {45}},\ \bibinfo {pages} {18} (\bibinfo {year} {2024})}\BibitemShut
  {NoStop}%
\bibitem [{\citenamefont {{Holland-Moritz}}\ \emph {et~al.}(2007)\citenamefont
  {{Holland-Moritz}}, \citenamefont {Heinen}, \citenamefont {Bellissent},\ and\
  \citenamefont {Schenk}}]{holland-moritzShortrangeOrderStable2007}%
  \BibitemOpen
  \bibfield  {author} {\bibinfo {author} {\bibfnamefont {D.}~\bibnamefont
  {{Holland-Moritz}}}, \bibinfo {author} {\bibfnamefont {O.}~\bibnamefont
  {Heinen}}, \bibinfo {author} {\bibfnamefont {R.}~\bibnamefont {Bellissent}},\
  and\ \bibinfo {author} {\bibfnamefont {T.}~\bibnamefont {Schenk}},\
  }\bibfield  {title} {\bibinfo {title} {Short-range order of stable and
  undercooled liquid titanium},\ }\href
  {https://doi.org/10.1016/j.msea.2005.12.093} {\bibfield  {journal} {\bibinfo
  {journal} {Materials Science and Engineering: A}\ }\bibinfo {series}
  {Proceedings of the 12th {{International Conference}} on {{Rapidly Quenched}}
  \& {{Metastable Materials}}},\ \textbf {\bibinfo {volume} {449--451}},\
  \bibinfo {pages} {42} (\bibinfo {year} {2007})}\BibitemShut {NoStop}%
\bibitem [{\citenamefont {Mayo}\ \emph {et~al.}(2013)\citenamefont {Mayo},
  \citenamefont {Yahel}, \citenamefont {Greenberg},\ and\ \citenamefont
  {Makov}}]{mayoShortRangeOrder2013}%
  \BibitemOpen
  \bibfield  {author} {\bibinfo {author} {\bibfnamefont {M.}~\bibnamefont
  {Mayo}}, \bibinfo {author} {\bibfnamefont {E.}~\bibnamefont {Yahel}},
  \bibinfo {author} {\bibfnamefont {Y.}~\bibnamefont {Greenberg}},\ and\
  \bibinfo {author} {\bibfnamefont {G.}~\bibnamefont {Makov}},\ }\bibfield
  {title} {\bibinfo {title} {Short range order in liquid pnictides},\ }\href
  {https://doi.org/10.1088/0953-8984/25/50/505102} {\bibfield  {journal}
  {\bibinfo  {journal} {Journal of Physics: Condensed Matter}\ }\textbf
  {\bibinfo {volume} {25}},\ \bibinfo {pages} {505102} (\bibinfo {year}
  {2013})}\BibitemShut {NoStop}%
\bibitem [{\citenamefont {Greenberg}\ \emph {et~al.}(2009)\citenamefont
  {Greenberg}, \citenamefont {Yahel}, \citenamefont {Caspi}, \citenamefont
  {Benmore}, \citenamefont {Beuneu}, \citenamefont {Dariel},\ and\
  \citenamefont {Makov}}]{greenbergEvidenceTemperaturedrivenStructural2009}%
  \BibitemOpen
  \bibfield  {author} {\bibinfo {author} {\bibfnamefont {Y.}~\bibnamefont
  {Greenberg}}, \bibinfo {author} {\bibfnamefont {E.}~\bibnamefont {Yahel}},
  \bibinfo {author} {\bibfnamefont {E.~N.}\ \bibnamefont {Caspi}}, \bibinfo
  {author} {\bibfnamefont {C.}~\bibnamefont {Benmore}}, \bibinfo {author}
  {\bibfnamefont {B.}~\bibnamefont {Beuneu}}, \bibinfo {author} {\bibfnamefont
  {M.~P.}\ \bibnamefont {Dariel}},\ and\ \bibinfo {author} {\bibfnamefont
  {G.}~\bibnamefont {Makov}},\ }\bibfield  {title} {\bibinfo {title} {Evidence
  for a temperature-driven structural transformation in liquid bismuth},\
  }\href {https://doi.org/10.1209/0295-5075/86/36004} {\bibfield  {journal}
  {\bibinfo  {journal} {Europhysics Letters}\ }\textbf {\bibinfo {volume}
  {86}},\ \bibinfo {pages} {36004} (\bibinfo {year} {2009})}\BibitemShut
  {NoStop}%
\bibitem [{\citenamefont {Itami}\ \emph {et~al.}(2003)\citenamefont {Itami},
  \citenamefont {Munejiri}, \citenamefont {Masaki}, \citenamefont {Aoki},
  \citenamefont {Ishii}, \citenamefont {Kamiyama}, \citenamefont {Senda},
  \citenamefont {Shimojo},\ and\ \citenamefont
  {Hoshino}}]{itamiStructureLiquidSn2003}%
  \BibitemOpen
  \bibfield  {author} {\bibinfo {author} {\bibfnamefont {T.}~\bibnamefont
  {Itami}}, \bibinfo {author} {\bibfnamefont {S.}~\bibnamefont {Munejiri}},
  \bibinfo {author} {\bibfnamefont {T.}~\bibnamefont {Masaki}}, \bibinfo
  {author} {\bibfnamefont {H.}~\bibnamefont {Aoki}}, \bibinfo {author}
  {\bibfnamefont {Y.}~\bibnamefont {Ishii}}, \bibinfo {author} {\bibfnamefont
  {T.}~\bibnamefont {Kamiyama}}, \bibinfo {author} {\bibfnamefont
  {Y.}~\bibnamefont {Senda}}, \bibinfo {author} {\bibfnamefont
  {F.}~\bibnamefont {Shimojo}},\ and\ \bibinfo {author} {\bibfnamefont
  {K.}~\bibnamefont {Hoshino}},\ }\bibfield  {title} {\bibinfo {title}
  {Structure of liquid {{Sn}} over a wide temperature range from neutron
  scattering experiments and first-principles molecular dynamics simulation:
  {{A}} comparison to liquid {{Pb}}},\ }\href
  {https://doi.org/10.1103/PhysRevB.67.064201} {\bibfield  {journal} {\bibinfo
  {journal} {Physical Review B}\ }\textbf {\bibinfo {volume} {67}},\ \bibinfo
  {pages} {064201} (\bibinfo {year} {2003})}\BibitemShut {NoStop}%
\bibitem [{\citenamefont {Greenberg}\ \emph {et~al.}(2010)\citenamefont
  {Greenberg}, \citenamefont {Yahel}, \citenamefont {Caspi}, \citenamefont
  {Beuneu}, \citenamefont {Dariel},\ and\ \citenamefont
  {Makov}}]{greenbergRelationMicroscopicStructure2010}%
  \BibitemOpen
  \bibfield  {author} {\bibinfo {author} {\bibfnamefont {Y.}~\bibnamefont
  {Greenberg}}, \bibinfo {author} {\bibfnamefont {E.}~\bibnamefont {Yahel}},
  \bibinfo {author} {\bibfnamefont {E.~N.}\ \bibnamefont {Caspi}}, \bibinfo
  {author} {\bibfnamefont {B.}~\bibnamefont {Beuneu}}, \bibinfo {author}
  {\bibfnamefont {M.~P.}\ \bibnamefont {Dariel}},\ and\ \bibinfo {author}
  {\bibfnamefont {G.}~\bibnamefont {Makov}},\ }\bibfield  {title} {\bibinfo
  {title} {On the relation between the microscopic structure and the sound
  velocity anomaly in elemental melts of groups {{IV}}, {{V}}, and {{VI}}},\
  }\href {https://doi.org/10.1063/1.3474997} {\bibfield  {journal} {\bibinfo
  {journal} {The Journal of Chemical Physics}\ }\textbf {\bibinfo {volume}
  {133}},\ \bibinfo {pages} {094506} (\bibinfo {year} {2010})}\BibitemShut
  {NoStop}%
\end{thebibliography}
\end{document}